\documentclass[10pt,journal,compsoc]{IEEEtran}

\usepackage{meta/lang}
\usepackage{color}
\usepackage{etoolbox}
\usepackage{listings}
\usepackage[T1]{fontenc}

\usepackage[scaled=0.8]{beramono}

\newcommand{\lstfontfamily}{\ttfamily}

\definecolor{darkviolet}{rgb}{0.5,0,0.4}
\definecolor{darkgreen}{rgb}{0,0.4,0.2} 
\definecolor{darkblue}{rgb}{0.1,0.1,0.9}
\definecolor{darkgrey}{rgb}{0.5,0.5,0.5}
\definecolor{lightblue}{rgb}{0.4,0.4,1}

\definecolor{stringColor}{rgb}{0.16,0.00,1.00}
\definecolor{annotationColor}{rgb}{0.39,0.39,0.39}
\definecolor{keywordColor}{rgb}{0.50,0.00,0.33}
\definecolor{commentColor}{rgb}{0.25,0.50,0.37}
\definecolor{javadocColor}{rgb}{0.25,0.37,0.75}
\definecolor{jTagColor}{rgb}{0.50,0.62,0.75}
\definecolor{eTagColor}{rgb}{0.50,0.62,0.75}
\definecolor{lineNumberColor}{rgb}{0.47,0.47,0.47}

\def\jTags{@author, @deprecated, @exception, @param, @return, @see, @serial, @serialData, @serialField, @since, @throws, @version}

\def\jAnnotations{
    classoffset=1,
    morekeywords={@Override, @Deperecated, @SuppressWarnings, @Retention, @Documented, @Target, @Inherited},
    keywordstyle=\color{annotationColor},
    classoffset=0
}

\def\eTags{FIXME, TODO, XXX}

\newrobustcmd{\markupJavadocs}[1]{%
\edef\mytok{\the\lst@token}%
\renewcommand*{\do}[1]{%
\ifdefstring{\mytok}{##1}%
{\color{jTagColor}\bfseries\listbreak}%
{}%
}%
{\color{javadocColor}%
\expandafter\docsvlist\expandafter{\jTags}%
\renewcommand*{\do}[1]{%
\ifdefstring{\mytok}{##1}%
{\color{eTagColor}\bfseries\listbreak}%
{}%
}%
\expandafter\docsvlist\expandafter{\eTags}%
#1}%
}%

\newrobustcmd{\markupComments}[1]{%
\edef\mytok{\the\lst@token}%
\renewcommand*{\do}[1]{%
\ifdefstring{\mytok}{##1}%
{\color{eTagColor}\bfseries\listbreak}%
{}%
}%
{\color{commentColor}%
\expandafter\docsvlist\expandafter{\eTags}#1}%
}%

\lstdefinestyle{eclipse}{
  basicstyle={\lstfontfamily},
  emphstyle=\bfseries,
  keywordstyle=\color{keywordColor}\bfseries,
  commentstyle=\markupComments,
  stringstyle=\color{stringColor},
  numberstyle=\color{lineNumberColor}\lstfontfamily,
  morecomment=[s][\markupJavadocs]{/**}{*/}, 
  showstringspaces=false,
  numbers=left,
  float=tp,
  floatplacement=tbp,
  belowskip=0em,
}

\lstdefinestyle{black}{
  basicstyle=\small\lstfontfamily,
  numbers=left,
  columns=fullflexible,
  breaklines=true,
  mathescape=true,
  escapechar=\#,
  tabsize=4,
  frame=lines,
  showstringspaces=false
}

\lstdefinestyle{seminar}{
  basicstyle=\small\ttfamily,
  numbers=left,
  breaklines=true,
  mathescape=true,
  escapechar=\#,
  tabsize=4,
  showstringspaces=false
}

\expandafter\lstset\expandafter{\jAnnotations}

\usepackage{hyperref}
\usepackage{mdframed}
\usepackage{lipsum}
\usepackage{amsmath}
\usepackage{amssymb}
\usepackage{amsfonts}
\usepackage[font=small]{caption}
\usepackage{ifpdf}
\usepackage[final]{graphicx}
\usepackage{todonotes}
\usepackage{booktabs} 
\usepackage{url}
\usepackage{amssymb}
\usepackage{pifont}
\usepackage{color}
\usepackage{tcolorbox}
\usepackage{balance}
\usepackage{makecell}
\usepackage{verbatim}
\usepackage{hyperref}
\usepackage{longtable}
\usepackage{rotating}
\usepackage{array}
\usepackage{tabularx}
\usepackage{adjustbox}
\usepackage{multirow}
\usepackage{siunitx}
\usepackage{pifont}
\usepackage{csquotes}
\usepackage{url}

\usepackage{xcolor}
\usepackage{fancybox}
\cornersize{.1} 

\newcommand{\mynewcontent}[2]{\ifnum#1<8{#2}\else{\textcolor{red}{#2}}\fi}

\providecommand{\inlinecode}[1]{\textcolor{black}{\texttt{#1}}}
\providecommand{\inlinekeyword}[1]{\textcolor{blue}{\texttt{#1}}}%

\definecolor{greylight}{RGB}{240,240,240}
\definecolor{greymedium}{RGB}{189,189,189}
\definecolor{greydark}{RGB}{99,99,99}

\usepackage{tcolorbox}
\tcbset{colback=greylight, colframe=greydark, arc=1pt, boxrule=0.75pt}

\newcommand{\java}{JAVA}
\newcommand{\evosuite}{EvoSuite}
\providecommand{\inlinecode}[1]{\textcolor{black}{\texttt{#1}}}
\providecommand{\inlinekeyword}[1]{\textcolor{blue}{\texttt{#1}}}%
\definecolor{rateHighBlue}{RGB}{75,146,246}
\newcommand{\precision}{$84.4\%$}
\newcommand{\recall}{$83\%$}
\newcommand{\fone}{$83.7\%$}
\newcommand{\ctinvo}{$\text{C3}_{\textbf{invo}}$}

\ifCLASSOPTIONcompsoc
  \usepackage[nocompress]{cite}
\else
  \usepackage{cite}
\fi
\ifCLASSINFOpdf
\else
\fi
\begin{document}
%
\title{An LLM-based Readability Measurement for Unit Tests' Context-aware Inputs}
%
%
%
%

\author{Zhichao~Zhou,
        Yutian~Tang,
        Yun~Lin,
        and~Jingzhu~He
\thanks{Z. Zhou and J. He are with School of Information Science and Technology, ShanghaiTech University, China}
\thanks{Y. Tang is with the University of Glasgow, United Kingdom}
\thanks{Y. Lin is with the Shanghai Jiao Tong University, China}
\thanks{J. He (hejzh1@shanghaitech.edu.cn) is the corresponding author.}}

%
%

\markboth{Journal of \LaTeX\ Class Files,~Vol.~14, No.~8, August~2015}%
{Shell \MakeLowercase{\textit{et al.}}: Bare Advanced Demo of IEEEtran.cls for IEEE Computer Society Journals}
%



\IEEEtitleabstractindextext{%
\begin{abstract}
Automated test generation tools (e.g., EvoSuite and Randoop) usually generate unit tests with higher code coverage than manual tests. However, the readability of automated tests is crucial for code comprehension and maintenance. The readability of unit tests involves many aspects. In this paper, we focus on test inputs. The central limitation of existing studies on input readability is that they focus on test codes alone without taking the tested source codes into consideration, making them either ignore different source codes' different readability requirements or require manual efforts to write readable inputs. However, we observe that the source codes specify the contexts that test inputs must satisfy. Based on such observation, we introduce the \underline{C}ontext \underline{C}onsistency \underline{C}riterion (a.k.a, C3), which is a readability measurement tool that leverages Large Language Models to extract primitive-type (including string-type) parameters' readability contexts from the source codes and checks whether test inputs are consistent with those contexts. We have also proposed EvoSuiteC3. It leverages C3's extracted contexts to help EvoSuite generate readable test inputs. We have evaluated C3's performance on $409$ \java{} classes and compared manual and automated tests' readability under C3 measurement. The results are two-fold. First, The Precision, Recall, and F1-Score of C3's mined readability contexts are \precision{}, \recall{}, and \fone{}, respectively. Second, under C3's measurement, the string-type input readability scores of EvoSuiteC3, ChatUniTest (an LLM-based test generation tool), manual tests, and two traditional tools (EvoSuite and Randoop) are $90\%$, $83\%$, $68\%$, $8\%$, and $8\%$, showing the traditional tools' inability in generating readable string-type inputs. We have conducted a survey based on the questionnaires collected from 30 programmers with varied backgrounds. The results reveal that when C3 identifies readable differences between tests, programmers tend to give similar opinions of the test's readability of C3.
\end{abstract}

\begin{IEEEkeywords}
  readability, test generation, large language models.
\end{IEEEkeywords}}

\maketitle

\IEEEdisplaynontitleabstractindextext

%
\IEEEpeerreviewmaketitle

\section{Introduction} \label{sec:intro}
\begin{figure*}[t]
    \centering
    \includegraphics[width=\textwidth]{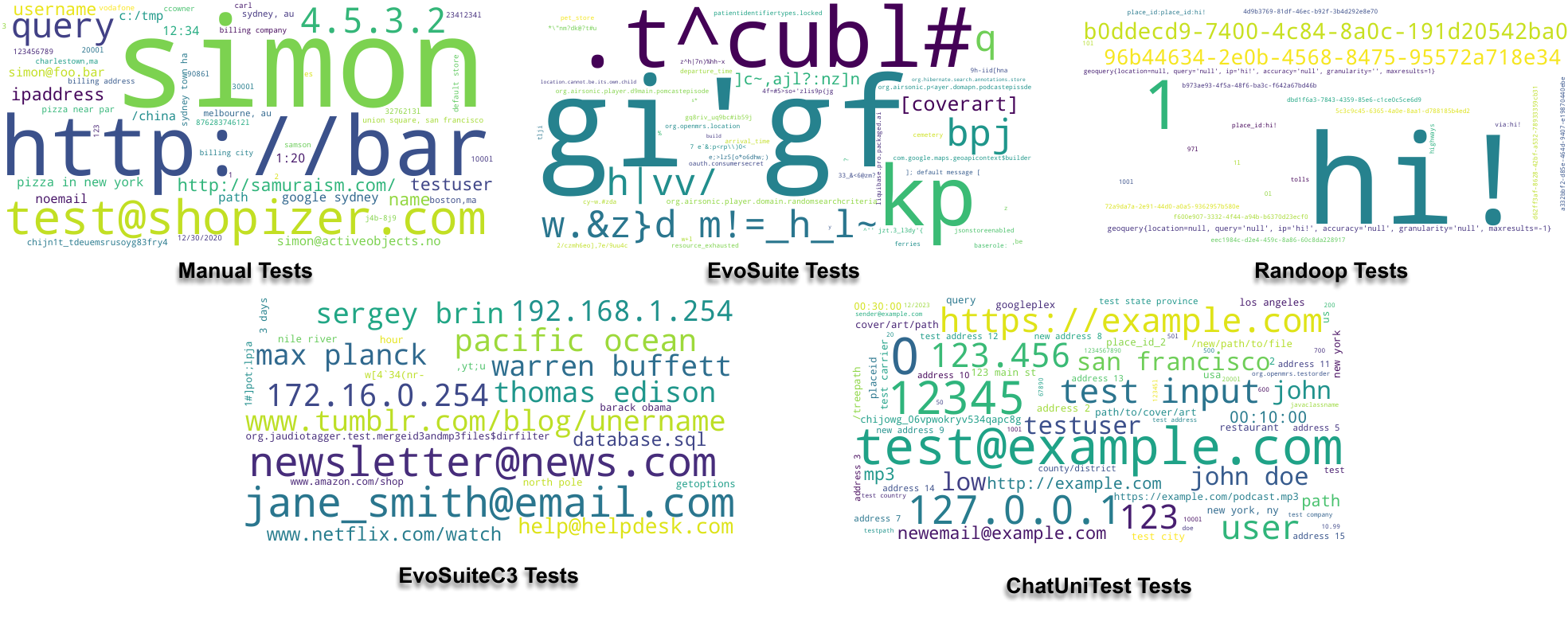}
    \caption{Overview of test inputs from manual and automated tests for $115$ parameters from $30$ \java{} classes (see Sec. ~\ref{sec:eval:compare}): Manual, EvoSuiteC3 (see Sec. ~\ref{fitness:c3}), and ChatUniTest~\cite{xie2023chatunitest} tests contain many real-world entities, such as person names, emails, URLs, and locations; EvoSuite~\cite{FraserEvoSuite} and Randoop~\cite{pacheco2007randoop} tests contain many random inputs, as well as \java{} class names, JSON strings, and UUIDs~\cite{JavaUUID} since these two tools tend to use predefined seeds and static or dynamic information of classes under test as test inputs}
    \label{fig:wordcloud}
\end{figure*}
Unit testing is a popular testing technique among developers. A unit test checks whether a method inside the source code behaves functionally correctly. Existing unit test generation tools (e.g., search-based tool EvoSuite ~\cite{FraserEvoSuite, PanichellaDynaMOSA} and feedback-directed random tool Randoop~\cite{pacheco2007randoop}) generate unit tests that are close to and even outperform the manually written ones in many coverage criteria~\cite{wang2021ML}, such as branch coverage.

We observe that automated tests' readability is crucial for programmers. For example, in a scenario where programmers use tools to obtain regression tests for two code versions~\cite{zhang2016isomorphic, salahirad2019choosing}, they must review a test failing at an assertion to locate and fix new bugs. It is time-consuming and takes lots of manual effort~\cite {mcminn2010reducing, afshan2013evolving, barr2014oracle}.
\begin{javacode}{A constructor from Airsonic~\cite{AirsonicUser} with five tests differing in inputs}{code:airsonic_user}
// method under test
public User(String username, String password, String email) {
    this.username = username; this.password = password;
    this.email = email;
}
public void testCreateUser() { // manual test
    User user = new User("Simon", "secret", "simon@example.org");
}
public void testConstructor() { // automated test by ChatUniTest
    User user = new User("testUser", "testPassword", "test@example.com");
}
public void test05() throws Throwable { // automated test by EvoSuite
    User user0 = new User("|x45e*3q4+", " [stream]", "");
}
public void test001() throws Throwable { // automated test by Randoop
    User user7 = new User("hi!", "", "");
}
public void test25() throws Throwable { // automated test by EvoSuiteC3
    User user0 = new User("Enrico Fermi", "guest", "careers@jobs.com");
}
\end{javacode}

The readability of unit tests involves many aspects, including but not limited to tests' method names~\cite{daka2017generating}, comments~\cite{panichella2016impact, roy2020deeptc}, documents~\cite{li2016automatically}, identifiers, and assertions~\cite{gay2023improving, daka2015modeling}. In this paper, we focus on test inputs. Many studies~\cite{afshan2013evolving, alsharif2019factors, almasi2017industrial, winkler2024investigating} and our survey (see Sec.~\ref{sec:eval:survey}) show that test inputs are crucial in readability. However, the random inputs generated by existing tools (EvoSuite and Randoop) cost understanding time. A former survey~\cite{afshan2013evolving} showed that compared to strings occurring in natural languages (e.g., \inlinecode{"mail"}), random strings (e.g., \inlinecode{"Qu5ua"}) require people at most $1.6$ times more comprehension time. To deal with this issue, Afshan et al.~\cite{afshan2013evolving} used a statistical model to predict the likelihood of the test input occurring in the real world. They formulated the likelihood as a secondary objective to guide test generation. The main issue with this approach is that its model ignores the source code when judging the input readability. However, different source codes have different requirements for input readability. For example, a parameter \inlinecode{email} calls for email addresses as readable inputs, while \inlinecode{city} requires city names. Deljouyi et al. \cite{deljouyi2023generating} proposed to extract understandable inputs from end-to-end (E2E) tests' execution trace logs since an E2E test aims to simulate a real user to execute an application’s entire workflow~\cite{leotta2024ai, augusto2020efficient} and may contain understandable inputs. However, this approach requires human effort to write E2E tests.
\subsection{A Motivating Example}
We leverage an example to demonstrate the lack of readability of automated tests and the challenges of existing studies. This example is a simplified constructor method under test (MUT) with a manual and two automated tests on Code~\ref{code:airsonic_user}. The MUT comes from an open-source project, Airsonic~\cite{AirsonicUser}. It has three parameters, which represent the user's name, password, and email, respectively. The manual test assigns the values \inlinecode{"Simon"}, \inlinecode{"secret"}, and \inlinecode{"simon@example.org"} to them, matching the contexts of the three parameters. The input values generated by an LLM-based test generation tool (ChatUniTest~\cite{xie2023chatunitest}) are \inlinecode{"testUser"}, \inlinecode{"testPassword"}, and \inlinecode{"test@example.com"}. They also match the contexts. Conversely, the input values from \evosuite{}~\cite{FraserEvoSuite} are \inlinecode{"|x45e*3q4+"}, \inlinecode{" [stream]"}, and \inlinecode{""}. These values make the automated test less readable than the manual one. In our survey (see Sec.~\ref{sec:eval:survey}), $22$ out of $30$ respondents prefer the manual test to EvoSuite's test regarding readability. The corresponding values from Randoop~\cite{pacheco2007randoop} are \inlinecode{"hi!"}, \inlinecode{""}, and \inlinecode{""}, posing the same issue. Such deficiency prevails in tests generated by traditional tools. Fig.~\ref{fig:wordcloud} aggregates the inputs from manual tests and tests generated by various testing tools for 30 classes (see Sec.~\ref{sec:eval:compare}). It shows that manual tests contain diversified real-world objects (e.g., person names, emails, and locations), while \evosuite{} and Randoop~\cite{pacheco2007randoop} generate a large number of random inputs. It could be worse for methods with multiple parameters. One interviewee gave us the following feedback:

\begin{quote}
``\textit{When there are many parameters, readers can not easily match parameters with values if a test uses a bunch of random inputs.}''
\end{quote}

The secondary objective complementing branch objectives proposed by Afshan et al.~\cite{afshan2013evolving} guides genetic algorithms~\cite{PanichellaDynaMOSA, FraserWhole} to generate inputs that are more likely to occur in the real language. However, this objective ignores the source code, thus ignoring the different input readability requirements of \inlinecode{username} and \inlinecode{email}. Hence, the optimized inputs may also not satisfy these readability requirements hidden in the source code. The approach of Deljouyi et al. \cite{deljouyi2023generating} to extract readable test inputs from manual-written E2E tests could not applied in this \java{} class under test without extra manual efforts since the Airsonic project does not have E2E tests. 

To cope with this readability issue, we have to solve two challenges:

\noindent$\bullet$\textbf{Challenge 1} is how to extract the readability requirements for the parameters from the source code. Such requirements may appear in many forms. For example, the forms denoting that a string-type parameter requires a person's name as its value include but are not limited to that (1) its parameter name is \inlinecode{userName}, (2) its parameter name is \inlinecode{value} but is followed by a comment \enquote{\textit{value is person name}}, and (3) its parameter name is \inlinecode{value}, but it is used by a SQL statement \inlinecode{"select * from user where name = \{value\}"}. Hence, it is challenging to identify such requirements using traditional techniques like regular expression matching or data-flow analysis~\cite{allen1976program}.

\noindent$\bullet$\textbf{Challenge 2} is how to improve existing traditional tools (e.g., EvoSuite) to fulfill such requirements after they are extracted. 
\subsection{Contributions} 
To solve the challenges shown in the motivating example, this paper proposes a novel concept called \textbf{readability context}, which refers to the requirements for test readability described in the source code (e.g., valid person names for the \inlinecode{username} parameter in Code~\ref{code:airsonic_user}). Based on the readability context, this paper proposes the Context Consistency Criterion (C3). First, C3 passes detailed information (including the source code and the comments) to a Large Language Model (LLM) and requests it to mine the readability contexts for the primitive-type parameters. We skip those complex-type parameters (e.g., object-types excluding string-type) since we only focus on the readability of literal input values (e.g., number and string literals) in this paper. Besides, we also pass few-shots~\cite{brown2020language} (i.e., pre-defined similar questions and their answers for LLM to quickly adapt to this mine-context task) to LLM. Second, C3 checks how well the inputs of the tests are consistent with readability contexts based on LLMs, NLP tools, and regex matching.

To sum up, this paper's contributions include:

\noindent$\bullet$ We observe that test inputs highly depend on the source codes. Based on the observation, we propose a novel concept, readability context, to represent the test readability requirements for the primitive-type variables of source codes. We design and implement an automatic tool, i.e., C3, to mine the readability contexts and leverage these contexts to measure the consistency of the tests' inputs with readability contexts (Sec.~\ref{sec:ccc}).

\noindent$\bullet$ We propose a variant of EvoSuite~\cite{FraserEvoSuite} (a search-based test generation tool), called EvoSuiteC3. EvoSuiteC3 leverages the readability contexts mined by C3 to help EvoSuite generate test inputs better in readability (Sec.~\ref{sec:eec}). The EvoSuiteC3 test in Code~\ref{code:airsonic_user} exemplifies readability improvements brought by EvoSuiteC3.

\noindent$\bullet$ We evaluate C3 on $107$ \java{} projects, including $409$ classes. C3 detects $1024$ ones with readability contexts from $4485$ primitive-type parameters. After a human review process, we confirm that C3's Precision, Recall, and F1-Score are \precision{}, \recall{}, and \fone{}, respectively (Sec.~\ref{sec:eval:accurate}).

\noindent$\bullet$ We compare manual tests, two traditional test generation tools (EvoSuite~\cite{FraserEvoSuite} and Randoop~\cite{pacheco2007randoop}), EvoSuiteC3, and an LLM-based tool (ChatUniTest~\cite{xie2023chatunitest}) on $30$ \java{} classes with $115$ parameters that have readability contexts. Fig.~\ref{fig:wordcloud} summarizes test inputs for these parameters and reveals that EvoSuiteC3 and ChatUniTest's tests perform closely to manual ones in input readability. Meanwhile, EvoSuite and Randoop largely fail to generate readable string-type inputs (Sec.~\ref{sec:eval:compare} and Sec.~\ref{sec:eval:evoc3}). Furthermore, we invite $30$ programmers with varied backgrounds to quantify the readability of manual and automated tests. The results show that programmers reach a consensus with C3 (Sec.~\ref{sec:eval:survey}).

The rest of this paper is organized as follows. Sec.~\ref{sec:related} discusses the related work. Sec.~\ref{sec:ccc} and~\ref{sec:eec} describe C3 and EvoSuiteC3's system design, respectively. Sec.~\ref{sec:eval} illustrates our evaluation. Sec.~\ref{sec:addcontext} shows how to extend C3. Sec.~\ref{sec:threats} presents the threats to validity. Sec.~\ref{sec:conclude} concludes the whole paper. Sec.~\ref{sec:data} provides the data availability.
\section{Related Work}\label{sec:related}
In this section, we discuss related studies.

\noindent\textbf{Traditional test generation.}
This paper uses two traditional test generation tools, Randoop~\cite{pacheco2007randoop} and EvoSuite~\cite{FraserEvoSuite}, to represent two methods. The first, random testing~\cite{pacheco2007randoop, PachecoFeedback, pacheco2005eclat, andrews2011genetic}, generates random test inputs to satisfy coverage criteria or expose faults in the software under test. Its lightweight nature allows for easy testing of large-scale systems, providing a solution to the scalability issues of static methods like model checking and symbolic execution~\cite{havelund2000model, braione2016jbse}. However, random testing has limitations. While its tests cover source lines, the random inputs may lack readability. Our evaluation~(Sec. \ref{sec:eval}) reveals that most string-type inputs generated by Randoop are meaningless (e.g., \inlinecode{"hi!"}), which confuse developers. Additionally, random testing without explicit guidance is inefficient in generating high-coverage tests. The second method, search-based testing~\cite{TonellaEvo2004, FraserWhole, PanichellaMOSA, PanichellaDynaMOSA, FraserEvoSuite, LinGraph}, addresses this by using coverage criteria (e.g., branch coverage) as fitness functions to guide genetic algorithms in generating test inputs. This approach may produce readable inputs if covering a criterion goal requires a readable input. For example, If a test covers the true branch of the branch condition \inlinecode{userName == "Peter"}, it contains a readable input (\inlinecode{"Peter"}) for a person's name. However, most readability requirements are not explicitly defined in real-world projects. Consequently, our experiments indicate that EvoSuite's readability is close to Randoop's (Sec.~\ref{sec:eval:compare}).

\noindent\textbf{Readability of traditional test generation.}
We divide existing studies on the readability of traditional test generation into two categories. The first focuses on generating test summaries. Daka et al.~\cite{daka2017generating} proposed summarizing a test's covered goals as its name. Panichella et al.~\cite{panichella2016impact} suggested a more thorough summary incorporating covered goals, the \java{} class being tested, and additional details to create comments. Roy et al.~\cite{roy2020deeptc} employed deep learning to name tests and internal variables and create comments. Li et al.~\cite{li2016automatically} proposed an approach to generate test documentation. These summary-based approaches ignore the concrete inputs, which are also regarded as necessary for readability (see Sec. \ref{sec:eval:survey}). The second category applies code readability studies to test readability. Based on Buse and Weimer's research on code readability~\cite{buse2009learning}, Daka et al.~\cite{daka2015modeling} gathered a dataset of tests along with their human-rated readability, used common code metrics as features (including identifier length, parentheses, test length, and casts), and added extra features for tests (e.g., assertions). They trained a linear regression model to improve the readability of automated unit tests. Grano et al.~\cite{Grano2018eval} leveraged the existing code readability model~\cite{Scalabrino2016ReadablityModel} to compare the readability of manual and automated tests. Gregory Gay~\cite{gay2023improving} proposed improving tests' readability by leveraging LLMs to refactor tests, such as dividing a test into small ones and formatting test code to satisfy common code style standards. This paper closely aligns with the studies of Afshan et al.~\cite{afshan2013evolving} and Deljouyi et al. \cite{deljouyi2023generating}. Afshan et al.~\cite{afshan2013evolving} incorporated a statistical language model into search-based test generation to predict the likelihood of the test input occurring in the real world and guide test generation as a secondary fitness function. Deljouyi et al. \cite{deljouyi2023generating} proposed to leverage end-to-end (E2E) tests to generate understandable unit test inputs since an E2E test simulates a real user to execute an application’s entire workflow and may contain understandable inputs. However, this approach requires human effort to write E2E tests. All of the above-mentioned studies ignore the source code, which plays a crucial role in determining the readability of the test code. C3 complements these studies. It mines the readability contexts of the primitive-type parameter from the source code and judges whether a test input is consistent with the context as a readability indicator. Our experiments show that C3 identifies the shortcomings of EvoSuite and Randoop in readability, and EvoSuiteC3 improves EvoSuite's readability.

\begin{figure*}[t]
    \centering
    \includegraphics[width=\textwidth]{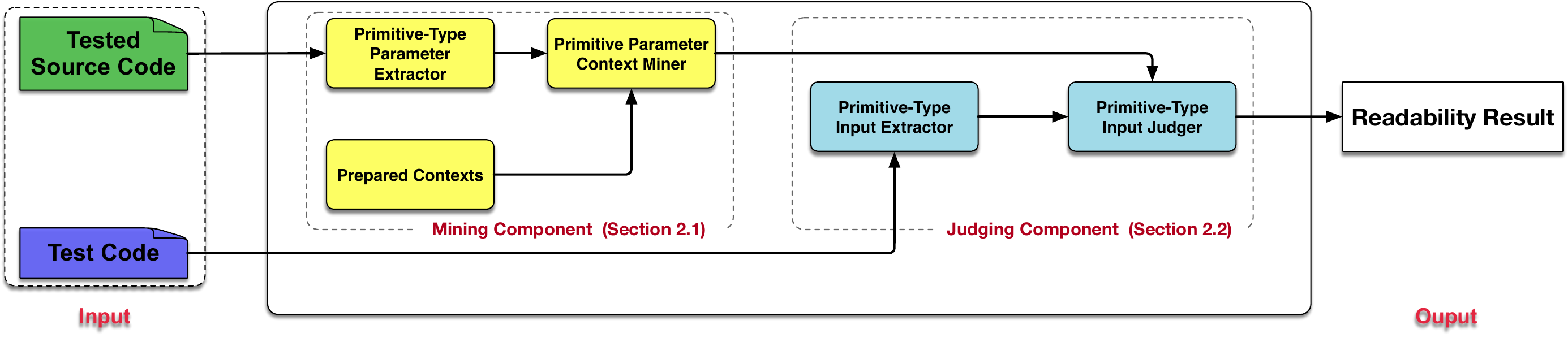}
    \caption{Overview of Context Consistency Criterion (C3)}
    \label{fig:overview}
\end{figure*}
\begin{table*}
\centering
\caption{Context Profile}
\label{tab:context_overview}
\scriptsize
\begin{tabular}{c|c|l|c}
\hline
\multirow{1}{*}{\textbf{Type}} & \multicolumn{1}{c|}{\textbf{Context Category}} & \multicolumn{1}{c|}{\textbf{Contexts}} & \multicolumn{1}{c}{\textbf{Judging Method}} \\ \cline{1-4}
 & Cyberspace &  EMAIL, FILE, PATH, PROGRAM, WEBSITE, URL, NETADDR & LLM + Regex matching \\ \cline{2-4}
  & Geo-Abstract &  CITY, ORGANIZATION, COUNTRY, GPE (e.g., \inlinecode{"UN"}), FAC (\inlinecode{"Route 66"}), LOCATION & LLM + CoreNLP~\cite{manning2014stanford} \\ \cline{2-4}
   & Temporal &  DATE (\inlinecode{"Sat 26 April"}), DURATION (\inlinecode{"15 years"}), TIMESET (\inlinecode{"Every Monday"}), TIME & LLM + CoreNLP \\ \cline{2-4}
    & Number &  PERCENT (\inlinecode{"20\%"}), ORDINAL (\inlinecode{"first"}), CARDINAL (\inlinecode{"two"}) & LLM + CoreNLP \\ \cline{2-4}
     & Personal Identifier &  PERSON (\inlinecode{"Simon"}), NORP (\inlinecode{"American"}) & LLM + CoreNLP \\ \cline{2-4}
\multirow{-6}{*}{String} & Finance & Money (\inlinecode{"2 euros"}) & LLM + CoreNLP \\ \cline{1-4}
\multirow{2}{*}{Number} & Base Formats & BINARY (\inlinecode{0b100}), HEXADECIMAL (\inlinecode{0xfe}), OCTAL (\inlinecode{007}) & Regex matching \\ \cline{2-4}
 & Other Formats & FIXEDLENGTH, LONGNUMBER, SCIENTIFIC (\inlinecode{1.2e20f}) & Rule + Regex matching \\ \hline
\end{tabular}
\end{table*}
\noindent\textbf{LLM-based test generation.}
Large language models (e.g., the GPT series~\cite{brown2020language}) have demonstrated capabilities across diverse domains, including unit test generation. Tang et al.~\cite{tang2023chatgpt} compared test suites generated by both LLM and EvoSuite in correctness, code coverage, complexity metrics~\cite{dantas2021readability}, and bug detection. Lemieux et al.~\cite{lemieux2023codamosa} suggested using LLMs to overcome plateaus in fitness functions during search-based test generation. Liu et al.~\cite{liu2023fill} proposed to leverage LLMs to improve GUI testing's page coverage. AthenaTest~\cite{tufano2020unit} employs a pre-trained BART model~\cite{lewis2019bart} for test generation, while ChatUniTest~\cite{xie2023chatunitest} is an automated tool that guides LLMs in generating and repairing tests. More studies are in the survey paper~\cite{wang2023software}. Given its ready-to-use nature and ability to continuously guide LLM in generating syntactically correct tests, this paper uses ChatUniTest as a baseline for comparison with EvoSuite and Randoop. Many LLM studies~\cite{tufano2020unit, xie2023chatunitest, yuan2024evaluating} assert that tests generated by LLMs are more readable than those produced by traditional tools, a conclusion reached through manual review or developer interviews. Our approach, C3, offers a quantitative metric for measuring test readability automatically. Our experimental results validate their findings that LLMs surpass EvoSuite and Randoop in generating readable tests. We also find that LLMs outperform developers as their tests are readable and achieve higher code coverage than manual tests.
\section{Context Consistency Criterion (C3)}\label{sec:ccc}
This section describes the design details of Context Consistency Criterion (C3). Fig.~\ref{fig:overview} shows its overview. Without losing any generality, we use \java{} as the target language to introduce C3, while C3 is not tied to any specific language. C3 takes the tested source code (e.g., a \java{} class or a method) and the test code as the input. C3's framework comprises two components, i.e., the Mining Component and the Judging Component. The Mining Component mines the readability contexts from the tested source code (Sec.~\ref{subsec:ccc_mine}). C3 extracts the primitive-type parameters from the source code. Then, for each parameter, C3 passes its text information (i.e., the source code and comments) to a Large Language Model (LLM) and lets the LLM choose a context from a group of prepared contexts (shown in Table~\ref{tab:context_overview}). The Judging Component measures the consistency of test inputs with the readability contexts (Sec.~\ref{subsec:ccc_judge}). C3 extracts the primitive-type inputs from the test. Then, it uses various tools (i.e., LLM, NLP tools, and regex matching) to check whether an input satisfies its context as the readability result.
\subsection{Mining Component}\label{subsec:ccc_mine}
The first step is to mine the contexts for each primitive-type parameter from the code under test. A simple method is to directly pass each parameter's text information (the source code and comments) to LLM and ask, \enquote{\textit{What context should this parameter's value match?}} However, such a method has two major flaws. First, an LLM may hallucinate~\cite{fan2023large} when facing an open-ended question, i.e., it may provide a fictitious response. For example, VHTest~\cite{huang2024visual} (an LLM benchmark) showed that, on average, $925$ of $1200$ open-ended questions induce four LLMs to hallucinate. Second, an LLM will return any possible result for such a question. Even if we regard this unpredictable result as correct, it is difficult to leverage it to judge tests' readability (Sec.~\ref{subsec:ccc_judge}) and improve existing tools (Sec.~\ref{fitness:c3}). Hence, we prepare a group of contexts (Table~\ref{tab:context_overview}) in advance so that we can restrict LLM to choosing only one of these contexts or answering no context. Our experiment (Sec.~\ref{subsec:ccc_mine}) showed that in a sample of $4485$ parameters, $1048$ ($23.4\%$) parameters were confirmed to be with our prepared contexts. Our design is extensible if end users add more specific contexts, and we provide a guide instruction (Sec.~\ref{sec:addcontext}) for users to add more contexts.

\noindent\textbf{Primitive-type parameter extractor.}
We leverage an open-source \java{} parser~\cite{JavaParser} to analyze the code and extract its parameters of primitive types, including \inlinekeyword{string}, \inlinekeyword{byte}, \inlinekeyword{short}, \inlinekeyword{int}, \inlinekeyword{long}, \inlinekeyword{float}, and \inlinekeyword{double}. Although \java{} regards \inlinekeyword{string} as an object type, the primitive types in this paper include it because \inlinekeyword{string} can be initialized by literal values like other primitive types. We omit \inlinekeyword{char} and \inlinekeyword{boolean} since the possible values of \inlinekeyword{char} are $0$ to $65536$ and \inlinekeyword{boolean}'s value is only either \inlinekeyword{true} or \inlinekeyword{false}. Hence, their range of concrete values is finite and small, posing no issues with readability. Besides, for a parameter, we extract its method's source and comments used to construct LLM's prompt. Note that we only use the method source code instead of the whole class code, thus significantly reducing the prompt size and decreasing the model's financial costs.

\noindent\textbf{Prepared contexts.}
We divide the concrete primitive types into two groups, i.e., String (including \inlinekeyword{string}) and Number (\inlinekeyword{byte}, \inlinekeyword{short}, \inlinekeyword{int}, \inlinekeyword{long}, \inlinekeyword{float}, and \inlinekeyword{double}). Then, we introduce their contexts by group.

\noindent\textit{(1) String.}
Named Entity Recognition (NER)~\cite{sang2003introduction, ritter2011named, manning2014stanford, wang2023gpt} is a natural language processing task that identifies key information in the text and classifies it into predefined categories (called named entities). A named entity refers to a real-world context-aware object, such as a person's name, organization, and email. We adopt named entities as the contexts for String. Referencing previous studies~\cite{manning2014stanford, wang2023gpt}, we prepare $23$ contexts shown in Table~\ref{tab:context_overview}, which are grouped into six context categories: Cyberspace ($7$), Geographic and Abstract Location ($6$), Temporal ($4$), Number (not Number type, but Number as String contexts) ($3$), Personal Identifier ($2$), and Finance ($1$). Note that \enquote{LOCATION} and \enquote{TIME} are two generic contexts. We use them to capture potential contexts in the source code that belong to Geo-Abstract Location and Temporal but do not match any other contexts in these two categories. For example, ``restaurant'' does not match any specific contexts  (e.g., \enquote{CITY} and \enquote{GPE}) of Geo-Abstract Location but is captured by \enquote{LOCATION}.

\noindent\textit{(2) Number.}
While named entities serve as String's contexts, we employ literal formats for Number's contexts, including base formats (e.g., \inlinecode{0b100}, \inlinecode{0xfe}) and Scientific Notation (e.g., \inlinecode{1.2e20f}). We add two more contexts, \enquote{FIXEDLENGTH} and \enquote{LONGNUMBER}. The former targets numbers of fixed lengths, such as credit card numbers, while the latter is designed for potentially overly long numbers.

It is worth noting that users can add extra contexts to meet their needs. Sec.~\ref{sec:addcontext} shows how to add new contexts.
\begin{figure}[t]
    \centering
    \includegraphics[width=0.46\textwidth]{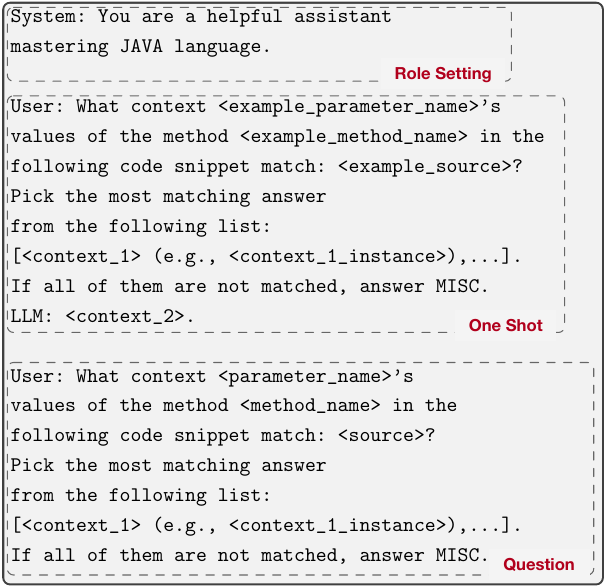}
    \caption{An example of context mining prompts for LLM}
    \label{fig:mining_example}
\end{figure}

\noindent\textbf{Primitive parameter context miner.}
Based on a primitive parameter's text information (the source code and comments) and the prepared contexts, we construct a prompt to ask an LLM which context the parameter matches. The prompt is created by following the few-shot technique~\cite{brown2020language}. Few-shot means that before asking a question, pass some similar questions and the prepared answers to an LLM in advance. This technique is proven to work well in many fields~\cite{brown2020language, ahmed2022few, ahmed2023better, liu2023large}. Fig.~\ref{fig:mining_example} shows our prompt example with one shot for one context category. The first part of Fig.~\ref{fig:mining_example} is to set the model role and control its behaviors. The second part is one shot, i.e., a question example with an answer. The third part is the question of interest. The first three placeholders store the parameter name, the method name, and the source code with comments. The following placeholders store each context and its instance (e.g., \enquote{PERSON} and \inlinecode{"Simon"}).

We start illustrating the prompt-constructing process by introducing two parameters for controlling the prompt size. In LLM, a token is a basic unit of text~\cite{brown2020language}, and an LLM has a token limit for a conversation. We use \textit{MaxToken} to represent the token size limit. Then, we use \textit{Remain} to keep the token space for LLM's response. This prompt-constructing process aims to add shots as many as possible. First, we calculate the \textit{RestToken} (to record the remaining token space) by $(\textit{MaxToken}- (\textit{Remain}+strToken(\textit{targetQuestion})))$, where \textit{strToken} computes the number of tokens within a string. Then, we add the role setting message to Messages and update \textit{RestToken}. Next, we iterate to add a shot and update \textit{RestToken} if there is enough token space. Finally, we add the targeting question. For the Number type, we have an extra operation. We observe that an important factor determining a Number parameter's context is the operators it uses. The \inlinecode{or} method on Code \ref{code:calc} presents an example. The \inlinecode{|} operator denotes that the values of \inlinecode{a} and \inlinecode{b} should be in binary format, i.e., their contexts are \enquote{BINARY}. Therefore, we emphasize this factor by appending a sentence to the end of the targeting question, i.e., \enquote{\texttt{The operators involved by this parameter are [op1, ...]}}. Meanwhile, two tests identical for compilers, \inlinecode{test1} and \inlinecode{test2}, vary in readability: The second is more readable since its inputs match the contexts. After generating messages, we pass them to LLM to get the result, which could be a concrete context or MISC. MISC means that LLM regards that no context matches this parameter.
\begin{javacode}{An \inlinecode{or} method with two tests differing in number formats}{code:calc}
// method under test
int or(int a, int b) { return a | b; }
// a test satisfying traditional criteria like line coverage
void test1() { assert 7 == or(4, 3); }
// a more readable test
void test2() { assert 0b111 == or(0b100, 0b011); }
\end{javacode}

\subsection{Judging Component}\label{subsec:ccc_judge}
The Judging Component aims to give a judgment for a test's input readability. It accepts parameters with contexts and a test as the inputs. For each parameter, it judges whether the test has an input passed to it and matches its context. We utilize various tools to judge the context matching as follows. First, we can organize a prompt for an LLM to judge inputs. Second, many NLP tools (e.g., CoreNLP~\cite{manning2014stanford} used in this paper) support the Named Entity Recognition (NER) task, i.e., extracting named entities from a piece of unstructured text. Since we use name entities as String's contexts, we further leverage CoreNLP's NER functionality to judge string inputs. Third, regex matching can judge inputs if one context has a corresponding regular expression (e.g., \enquote{EMAIL}). We first introduce each of them. Then, we describe combining them to complete the judging process.
\begin{figure}[t]
    \centering
    \includegraphics[width=0.46\textwidth]{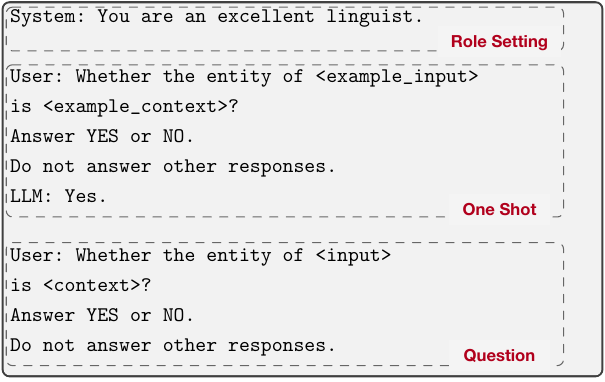}
    \caption{An example of context matching prompts for LLM}
    \label{fig:matching_example}
\end{figure}

\noindent\textbf{Tools.} We introduce the tools to perform context matching as follows.

\noindent\textit{(1) LLM.}
Fig.~\ref{fig:matching_example} shows an example of asking LLM to check whether an input matches the context. We prepare a shot for each context category like in the Mining Component (see Sec.~\ref{subsec:ccc_mine}). The prompt-constructing process is similar to the process of constructing mining prompts, too.

\noindent\textit{(2) CoreNLP.}
We use CoreNLP (a traditional NLP tool) to verify the consistency of a character string to the context via its NER functionality. CoreNLP supports all context categories of String~\cite{manning2014stanford}, except Cyberspace (see Table~\ref{tab:context_overview}). Its advantage over LLM is that it is financially cost-free.

\noindent\textit{(3) Regex matching.}
Regex matching works for String's Cyberspace and Number's all categories. For example, we use \inlinecode{"\textasciicircum 0[bB][01]+\$"} to check whether a number is binary.

\noindent\textbf{Primitive-type input extractor.}
We use a \java{} parser~\cite{JavaParser} to extract the test's inputs whose parameter has a context. Then, we run the judger on each input. Note that we only take three input types into account: the literal values, the primitive-type variables defined with a literal in the test method, and the test's \inlinecode{final} \inlinecode{static} fields. This choice ignores many input sources, such as another class's field. However, according to our survey (see Sec.~\ref{sec:eval:survey}), programmers prefer straightforward test statements. For example, pass-by-value is preferable to other ways when calling the method under test. Hence, this choice has a tiny impact on readability judgment.

\noindent\textbf{Primitive-type input judger.}
We use different tool combinations for context categories (shown in Table~\ref{tab:context_overview}'s last column). We prefer to use multiple tools in combination instead of a single tool to judge inputs because we want to reduce the misjudgment of readable inputs.

\noindent\textit{(1) LLM + Regex matching.}
We use this combination to check String's Cyberspace category (see Table~\ref{tab:context_overview}). Regex matching is enough to check whether an input is valid for these contexts (e.g., \enquote{EMAIL}). However, in realistic development, programmers may use an invalid but readable input, such as \inlinecode{"testEmail"} for \enquote{EMAIL}. Hence, we complement LLM to make the judge results more elastic than they are judged by regex matching only. We regard an input as readable as long as one tool gives a positive result.

\noindent\textit{(2) LLM + CoreNLP.}
This combination checks all other String's categories. Like the previous combination, an input is readable if at least one tool says yes, reducing the risk of ignored readable inputs.

\noindent\textit{(3) Regex matching.}
We use regex matching to check Number's Base Formats, such as \inlinecode{"\textasciicircum 0[bB][01]+\$"} for \enquote{BINARY}.

\noindent\textit{(4) Rule + Regex matching.}
This combination checks the Number's Other Formats. Before leveraging regex matching, we add a parameter called \textit{BearableLength}. For this category, if an input's length is shorter than \textit{BearableLength}, we regard it as readable. This paper sets \textit{BearableLength} to $9$, and we will discuss it in Sec.~\ref{sec:threats}. Otherwise, we use regex matching to check the inputs: the requirement for \enquote{LONGNUMBER} is that an underscore must appear after at most every three consecutive digits (e.g., \inlinecode{11\_321\_22\_745}); for \enquote{FIXEDLENGTH}, the requirement is that an underscore must exist and appear after a fixed number of consecutive digits (e.g., \inlinecode{1111\_1111\_1111\_1111}); for \enquote{SCIENTIFIC}, we checks whether the input contains an \inlinecode{e} or \inlinecode{E}. We assume the test code is grammatically correct, so as long as a number contains such symbols, it satisfies \enquote{SCIENTIFIC}.
\section{EvoSuite Enhanced by C3 (EvoSuiteC3)}\label{sec:eec}
The readability contexts mined by C3's Mining Component (Sec.~\ref{subsec:ccc_mine}) are not only used by C3's Judging Component (Sec.~\ref{subsec:ccc_judge}) to judge tests but also help automated tools generate tests with better input readability. We aim to help automated tools generate tests with more readable inputs and consistency in other perspectives (such as coverage) with the original tools. In this paper, we take EvoSuite as an example and propose EvoSuiteC3, i.e., EvoSuite~\cite{FraserEvoSuite} enhanced by C3. We choose EvoSuite instead of Randoop because EvoSuite is a search-based test generation tool. It formulates coverage metrics (e.g., Branch Coverage and Weak Mutation~\cite{Rojas2015CombiningMC, FraserMutation}) as fitness functions. It uses them to guide genetic algorithms~\cite{FraserWhole, PanichellaMOSA, PanichellaDynaMOSA} to evolve tests until all fitness functions reach optimum or the search budget is out. Hence, we integrate C3 into EvoSuite by formulating C3's contexts as the genetic algorithms' fitness functions. However, Randoop generates a test by repeating randomly calling the methods of the \java{} class under test until some predefined contract is violated~\cite{pacheco2007randoop}, which means that one test generated by Randoop aims to capture one behavior (i.e., a violated contract) of the class under test. Hence, it is difficult to add additional testing objectives to Randoop. 
\begin{figure}[t!]
    \centering
    \includegraphics[width=0.45\textwidth]{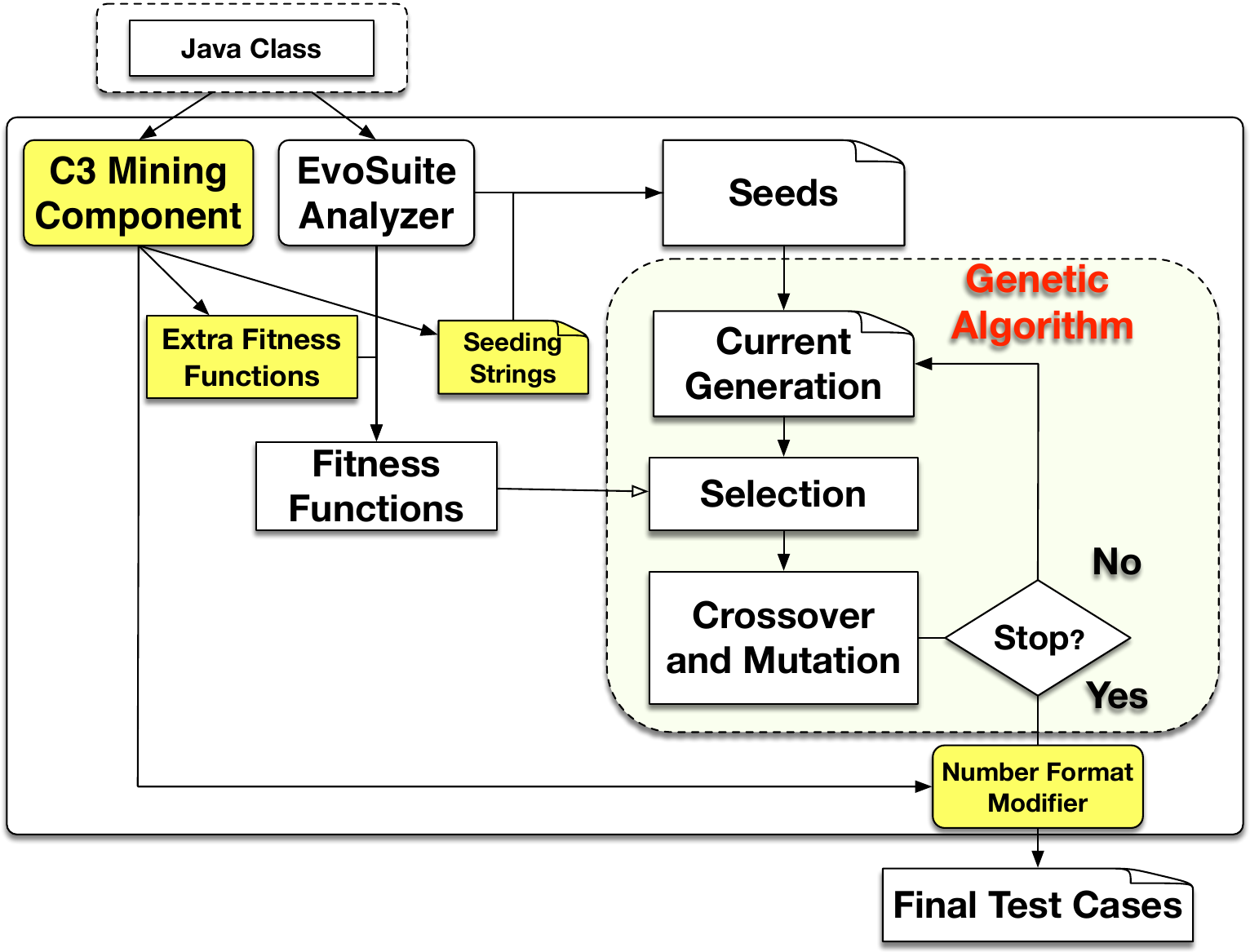}
    \caption{Overview of EvoSuiteC3}
    \label{fig:evosuitec3}
\end{figure}

Fig.~\ref{fig:evosuitec3} shows EvoSuiteC3's overview. Generally, we formulate C3's string-type contexts as fitness functions to guide EvoSuite's genetic algorithms to generate readable string-type inputs during the test generation process and modify the formats of numbers to satisfy their parameters' contexts after the test generation process. Specifically, for each string context, we provide a group of string seeds (e.g., \inlinecode{"Beijing"} and \inlinecode{"London"} for \enquote{CITY}) for EvoSuite. These seeds are randomly selected by EvoSuite to be directly inserted into tests. Secondly, we provide extra fitness functions to guide EvoSuite's genetic algorithm in generating tests with string values that satisfy contexts. Unlike string contexts, number contexts focus on the numbers' formats instead of values. Hence, we modify the formats of numbers according to contexts after generating tests without any operations in the test generation process.

To demonstrate the fitness function for string contexts, we first introduce several auxiliary functions: $getSeeds(context)$ gets prepared string seeds by a specific context. $isSatisfy(value, context)$ checks whether a string value satisfies a specific context by invoking C3's Judging Component (Sec.~\ref{subsec:ccc_judge}). It returns $1$ if the value satisfies the context; otherwise, it returns $0$. Note that EvoSuiteC3 forces C3 to only use the regex matcher and CoreNLP without LLM for judging (see Sec.~\ref{subsec:ccc_judge}) so that EvoSuiteC3 does not waste EvoSuite's time budget due to network calls and avoids model costs. $stringDistance(value1, value2)$ measures the distance (ranging from $0$ to $1$) between two string values. It returns $0$ if they are identical. We use Jaro-Winkler Distance~\cite{winkler1990string} to implement this function.

Based on these auxiliary functions, we define the readability fitness function $f_{c3}(value, context)$ as follows:
\begin{equation}
\label{fitness:c3}
    \left\{
\begin{array}{lcl}
    0 & & {isSatisfy(value,} \\
    & & {context)=1,} \\
    minDistance(value, context) & & {\textnormal{otherwise,}} \\
\end{array}\right.
\end{equation}
where $minDistance(value, context)$ is a function defined as $\min{\{stringDistance(value, v)|v \in getSeeds(context)\}}$.

Besides, a method may have multiple parameters with a context (e.g., the constructor in Code~\ref{code:airsonic_user}). To guide Evosuite in generating a method invocation satisfying multiple contexts, we define an extra fitness function $f_{c3invo}(invocation)$ as:
\begin{equation}
\label{fitness:c3invo}
 \max{\{f_{c3}(v, c)| (v,c) \in getVCPairs(invocation) \}},
\end{equation}
where $getVCPairs$ returns an invocation's all value-context pairs. This function returns the max return values of all the target method's $f_{c3}$ fitness functions. Hence, it returns $0$ if only all $f_{c3}$ fitness functions return $0$, i.e., all contexts are satisfied.

Based on the above definitions, EvoSuiteC3 extracts all $f_{c3}$ and $f_{c3invo}$ functions for parameters with contexts and methods containing such parameters from the \java{} class under test. It combines them with the existing fitness functions of EvoSuite to guide test generation.
\section{Evaluation}\label{sec:eval}
\begin{figure*}[t]
    \centering
    \includegraphics[width=\textwidth]{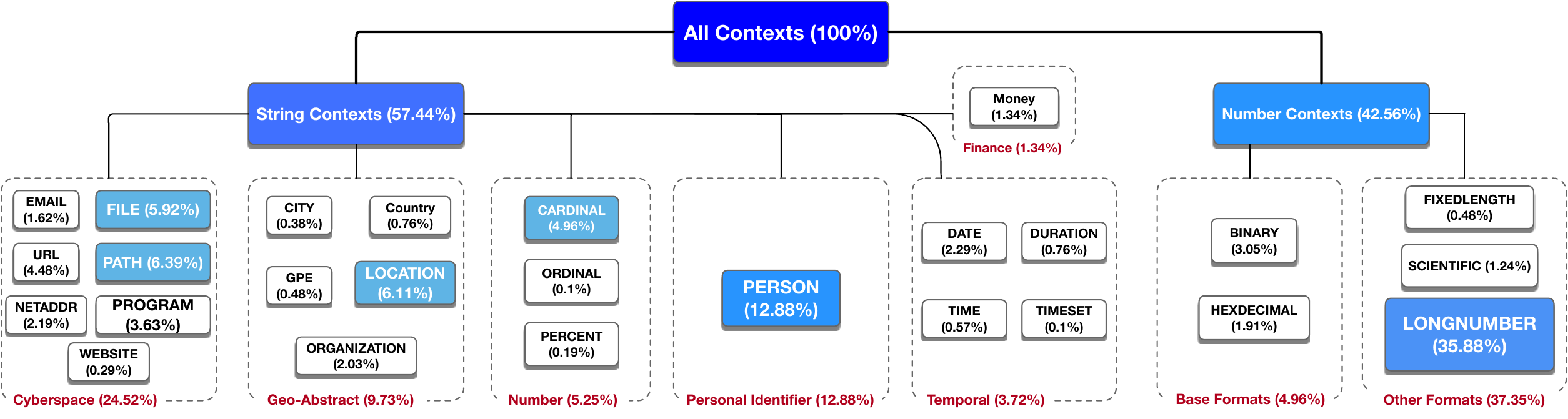}
    \caption{Readability contexts' proportion}
    \label{fig:rc}
\end{figure*}
The evaluation mainly assesses C3's performance, i.e., false positive and negative rates, in mining readability contexts from real projects. Additionally, we aim to explore the differences between manual and automated tests under C3 measurement and the improvement brought by EvoSuiteC3 to EvoSuite. Moreover, we investigate whether programmers agree with C3. The research questions are as follows:

\noindent$\bullet$\textbf{RQ1:}
(C3’s Performance) What are the false positive and negative rates of C3?

\noindent$\bullet$\textbf{RQ2:}
(Readability of Manual and Automated Tests) What differentiates manual from automated tests in C3 measurement?

\noindent$\bullet$\textbf{RQ3:}
(EvoSuiteC3's Improvement) How much improvement does EvoSuiteC3 bring to EvoSuite?

\noindent$\bullet$\textbf{RQ4:}
(Programmers’ Readability Measurements) Whether programmers agree with C3?
\subsection{RQ1: C3's Performance}\label{sec:eval:accurate}

This RQ aims to answer what the false positive and negative rates of C3 are.

\subsubsection{Methodology}
\begin{table}[htbp]
    \centering
    \small
    \caption{\java{} projects used in our study group by category.}
    \begin{tabular}{c|c|c}
    \hline
        \textbf{Category} & \textbf{Projects} & \textbf{Classes}  \\ \hline
         Frameworks and Libraries & 48 & 173  \\ 
        Utilities and Tools	 & 40 & 103  \\ 
        Content Management Systems & 13 & 125  \\ 
        Games and Entertainment & 6 & 8  \\ \hline
        Total & 107 & 409  \\ \hline
    \end{tabular}
    \label{tab:projects}
\end{table}
\noindent\textit{(1) Subjects.}
Table~\ref{tab:projects} presents the real projects used in this RQ. These projects come from two sources: (1) We randomly select $254$ classes from $100$ projects within the DynaMOSA benchmark~\cite{PanichellaDynaMOSA, CamposCombineDefect}; (2) To enrich the project diversity and obtain classes with manual tests for subsequent RQs, we add $155$ classes from seven open-source projects, namely, Airsonic~\cite{Airsonic}, BroadleafCommerce~\cite{BroadleafCommerce}, GoogleMapService~\cite{GoogleMapService}, OpenMRS~\cite{Openmrs}, OpenRefine~\cite{OpenRefine}, PetClinic~\cite{PetClinic}, and Shopizer~\cite{Shopizer}. In summary, these classes contain $4485$ primitive-type parameters.

\noindent\textit{(2) Procedure.}
We employ GPT-4-Turbo~\cite{GPTModels} to run with C3's mining component (Sec.~\ref{subsec:ccc_mine}). Specifically, C3 uses GPT-4-Turbo to analyze every primitive-type parameter and further extract readability contexts. The default GPT-4-Turbo settings are used when invoking its APIs. Two parameters' concrete values of C3's mining component (see Sec.~\ref{subsec:ccc_mine}) are $4096$ (\textit{MaxToken}) and $20$ (\textit{Remain}). Although multiple executions of this procedure could mitigate the effects of the LLM's randomness, C3 only uses the result of one run due to model financial costs. Then, we conduct a manual review of C3's results. Following the previous study~\cite{wang2021exploratory}, two authors independently examined each parameter, its associated \java{} source code, and comments to check whether the readability context is correct (True Positive), wrong (False Positive), or missed (False Negative). They then compared the label results and found a few differences. In the end, they discussed and resolved the differences. Note that a context may be both False Positive and False Negative, as C3 might incorrectly assign a context to a parameter with a correct readability context. In such instances, they only label these cases as False Positive.

\noindent\textit{(3) Data analysis.}
Firstly, we summarize the results as a Confusion Matrix~\cite{powers2020evaluation}. Secondly, we calculate the Precision, Recall, and F1-Score~\cite{powers2020evaluation}. Thirdly, we group C3's results by project categories to show the distribution of readability contexts. Finally, we compute the proportion of each context type.

\subsubsection{Results}
\begin{table}[htbp]
\centering
\caption{C3 accuracy summary: 240 parameters}
\begin{tabular}{c|c|c}
\hline
 & \textbf{C3 Positive} & \textbf{C3 Negative} \\ \hline
\textbf{Actual Positive} & 864 & 177 \\ \hline
\textbf{Actual Negative} & 160 & 3284 \\ \hline
\end{tabular}
\label{tab:rq2_cm}
\end{table}
\begin{table}[htbp]
    \centering
    \small
    \caption{Readability contexts group by project category}
    \begin{tabular}{c|c|c}
    \hline
        \textbf{Category} & \multicolumn{2}{c}{\textbf{Primitive-Type Parameters}}  \\ \cline{2-3}
        & \textbf{With contexts} & \textbf{Total} \\ \hline
         Frameworks and Libraries & 604 ($21.3\%$) & 2832    \\ 
         Utilities and Tools	 & 212 ($25.9\%$) & 820  \\ 
        Content Management Systems & 226 ($28.1\%$) & 804  \\ 
        Games and Entertainment & 6 ($20.7\%$) & 29  \\ \hline
        Total & 1048 ($23.4\%$) & 4485  \\ \hline
    \end{tabular}
    \label{tab:project_contexts}
\end{table}
\begin{table*}[htbp]
    \centering
    \caption{Projects used in RQ2-4}
    \begin{tabular}{c|c|c|c|c|c|c|c|c}
    \hline
        \textbf{Project} & \textbf{Category} & \textbf{Classes} & \multicolumn{3}{c|}{\textbf{Contexts}} & \multicolumn{3}{c}{\textbf{True Positive Contexts}} \\ \cline{4-9}
        & & & \textbf{Total} & \textbf{String} & \textbf{Number} & \textbf{Total} & \textbf{String} & \textbf{Number} \\ \hline
         Airsonic~\cite{Airsonic} & Utilities and Tools & 9 & 42 & 26 & 16 & 37 & 21 & 16 \\ 
          GoogleMapService~\cite{GoogleMapService}  & Frameworks and Libraries & 6 & 20 & 20 & 0 & 19 & 19 & 0 \\ 
         OpenMRS~\cite{Openmrs}  & Content Management Systems & 3 & 31 & 31 & 0 & 26 & 26 & 0 \\ 
         Shopizer~\cite{Shopizer} & Content Management Systems & 8 & 20 & 19 & 1 & 19 & 18 & 1 \\ 
          Twitter4J~\cite{PanichellaDynaMOSA} & Frameworks and Libraries & 4 & 15 & 8 & 7 & 14 & 7 & 7 \\ \hline
            Total & - & 30 & 128 & 104 & 24 & 115 & 91 & 24 \\ \hline
    \end{tabular}
    \label{tab:projects_rq2}
\end{table*}
\noindent\textit{(1) Confusion Matrix.}
Table~\ref{tab:rq2_cm} shows the True Positives (TPs), False Positives (FPs), True Negatives (TNs), and False Negatives (FNs).

\noindent\textit{(2) Metrics.}
{\footnotesize
\begin{equation}
    Precision = \frac{TP}{TP+FP}=\frac{864}{864+160}=84.4\%.\label{m:precision}
\end{equation}
\begin{equation}
    Recall = \frac{TP}{TP+FN}=\frac{864}{864+177}=83\%.\label{m:recall}
\end{equation}
\begin{equation}
    F1 = 2 \times \frac{Precision \times Recall}{Precision + Recall}=2 \times \frac{84.4\%\times 83\%}{84.4\%+83\%}= 83.7\%.\label{m:f1}
\end{equation}}
\noindent\textit{(3) Context distribution.}
After manually checking C3's results, we confirmed that $1048$ ($23.4\%$) out of $4485$ primitive-type parameters have readability contexts. Table~\ref{tab:project_contexts} presents the distribution of readability contexts across each group.

\noindent\textit{(4) Context proportion.}
Fig.~\ref{fig:rc} shows the proportions of $1048$ parameters' readability contexts. Overall, string-type readability contexts constitute $57.44\%$, while number-type contexts represent $42.56\%$. The leading two context categories are \enquote{Other Formats} ($37.35\%$) and \enquote{Cyberspace} ($24.52\%$). The top six contexts are \enquote{LONGNUMBER} ($35.88\%$), \enquote{PERSON} ($12.88\%$), \enquote{PATH} ($6.39\%$), \enquote{LOCATION} ($6.11\%$), \enquote{FILE} ($5.92\%$) and \enquote{CARDINAL} ($4.96\%$).
\subsubsection{Discussion}
Table~\ref{tab:project_contexts} shows that all project categories have a similar proportion of readability contexts ($20\%$ to $30\%$). Second, \enquote{LONGNUMBER} ($35.88\%$) is the most common context. After a manual review, we find that GPT-4-Turbo tends to associate a parameter with \enquote{LONGNUMBER} if its name ends with ``id'' (e.g., ``id'' and ``userId'') and concrete type is \inlinekeyword{long}. GPT-4-Turbo might do this because such parameters are typically used for large-value database keys. Meanwhile, such parameters are common in data-manipulating projects, leading to a high proportion of this context.

The statistical data shows C3's ability to correctly identify positive instances while minimizing false positive cases. There are still $177$ FNs and $160$ FPs. We classify the analyzed reasons as follows.

\noindent\textit{(1) LLM's randomness.}
We find $154$ ($87\%$) FNs and $49$ ($31\%$) FPs whose projects contain many parameters with similar names, comments, and meanings that are accurately annotated with context, except for them. We assume this is due to LLM's randomness, which could be mitigated by calling LLM multiple times. We randomly select $30$ false cases as a sample. After calling LLM three times, $20$'s readability contexts are correctly extracted at least one time. After calling LLM five times, the correct cases increase to $24$.

\noindent\textit{(2) LLM's insufficient understanding.}
We find that $28$ ($13\%$) FNs and $102$ ($64\%$) FPs are due to LLM's insufficient understanding. One example is a class \inlinecode{Playlist} from Airsonic~\cite{Airsonic}. Its one parameter, \textbf{\inlinecode{name}} (the music playlist's name), is wrongly recognized as \enquote{PERSON}. We assume that these cases are due to LLM's insufficient understanding, which could be solved by the LLM with better understanding in the future. On the other hand, our mine-context task can be used as a benchmark to compare different LLMs.

Another issue undermining C3 is that our contexts are incomplete. One example is a class \inlinecode{Role} from OpenMRS~\cite{Openmrs} (a medical record system). It has one string parameter, \textbf{\inlinecode{role}}, whose expected readable values include ``Medical Student'', ``Data Manager'', ``Data Assistant'', etc. But it is not recognized since we do not define \enquote{Role} context. We do not count these instances as FN, as they fall outside C3's contexts and exceed its capabilities. To cope with this issue, we allow C3's users to integrate new contexts into C3. Sec.~\ref{sec:addcontext} shows how to add contexts.
\begin{tcolorbox}[title=Answer to RQ1,boxrule=1pt,boxsep=1pt,left=2pt,right=2pt,top=2pt,bottom=2pt]
In a sample of $4485$ primitive-type parameters, C3 detects 1024 ones with readability contexts. After a manual review process, we confirm that its Precision, Recall, and F1-Score are \precision{}, \recall{}, and \fone{}, respectively.
\end{tcolorbox}
\begin{table*}
\centering
\caption{Judge summary group by project with true positive contexts (Text marked with blue background when Read.\% >70\%}
\label{tab:judge_tp}
\scriptsize
\begin{adjustbox}{width=\textwidth,center}
\begin{tabularx}{\textwidth}{c|c|>{\raggedleft\arraybackslash}X|>{\raggedleft\arraybackslash}X|>{\raggedleft\arraybackslash}X|>{\raggedleft\arraybackslash}X|>{\raggedleft\arraybackslash}X|>{\raggedleft\arraybackslash}X|>{\raggedleft\arraybackslash}X|>{\raggedleft\arraybackslash}X|>{\raggedleft\arraybackslash}X|>{\raggedleft\arraybackslash}X}
 \hline
\multicolumn{2}{c|}{\multirow{2}{*}{\textbf{Project}}} & \multicolumn{2}{c|}{\textbf{Manual}} & \multicolumn{2}{c|}{\textbf{ChatUniTest}} & \multicolumn{2}{c|}{\textbf{EvoSuite}} & \multicolumn{2}{c|}{\textbf{Randoop}} & \multicolumn{2}{c}{\textbf{EvoSuiteC3}} \\ \cline{3-12} 
\multicolumn{2}{c|}{} & \textbf{Cover. (\%)} & \textbf{Read. (\%)} & \textbf{Cover. (\%)} & \textbf{Read. (\%)} & \textbf{Cover. (\%)} & \textbf{Read. (\%)} & \textbf{Cover. (\%)} & \textbf{Read. (\%)} & \textbf{Cover. (\%)} & \textbf{Read. (\%)} \\ \hline

\multirow{2}{*}{Airsonic} & string & 15 (71\%) & 8 (53\%) & 21 (100\%) & \colorbox{rateHighBlue}{\textbf{\textcolor{white}{17 (81\%)}}} & 19 (90\%) & 1 (5\%) & 19 (90\%) & 4 (21\%) & 21 (100\%) & \colorbox{rateHighBlue}{\textbf{\textcolor{white}{20 (95\%)}}} \\
 & number & 10 (62\%) & \colorbox{rateHighBlue}{\textbf{\textcolor{white}{9 (90\%)}}} & 15 (94\%) & \colorbox{rateHighBlue}{\textbf{\textcolor{white}{15 (100\%)}}} & 11 (69\%) & \colorbox{rateHighBlue}{\textbf{\textcolor{white}{11 (100\%)}}} & 8 (50\%) & \colorbox{rateHighBlue}{\textbf{\textcolor{white}{8 (100\%)}}} & 10 (62\%) & \colorbox{rateHighBlue}{\textbf{\textcolor{white}{10 (100\%)}}} \\ \hline
 
GoogleMap & string & 12 (63\%) & \colorbox{rateHighBlue}{\textbf{\textcolor{white}{10 (83\%)}}} & 17 (89\%) & 9 (53\%) & 16 (84\%) & 2 (12\%) & 16 (84\%) & 0 (0\%) & 17 (89\%) & \colorbox{rateHighBlue}{\textbf{\textcolor{white}{13 (76\%)}}} \\
Service & number & - & - & - & - & - & - & - & - & - & - \\ \hline
 
\multirow{2}{*}{OpenMRS} & string & 0 (0\%) & - & 26 (100\%) & \colorbox{rateHighBlue}{\textbf{\textcolor{white}{25 (96\%)}}} & 26 (100\%) & 2 (8\%) & 20 (77\%) & 1 (5\%) & 26 (100\%) & \colorbox{rateHighBlue}{\textbf{\textcolor{white}{22 (85\%)}}} \\
 & number & - & - & - & - & - & - & - & - & - & - \\ \hline
 
\multirow{2}{*}{Shopizer} & string & 13 (72\%) & 9 (69\%) & 18 (100\%) & \colorbox{rateHighBlue}{\textbf{\textcolor{white}{17 (94\%)}}} & 18 (100\%) & 0 (0\%) & 18 (100\%) & 1 (6\%) & 18 (100\%) & \colorbox{rateHighBlue}{\textbf{\textcolor{white}{18 (100\%)}}} \\
 & number & 0 (0\%) & - & 0 (0\%) & - & 0 (0\%) & - & 0 (0\%) & - & 0 (0\%) & - \\ \hline
 
\multirow{2}{*}{Twitter4J} & string & 4 (57\%) & \colorbox{rateHighBlue}{\textbf{\textcolor{white}{3 (75\%)}}} & 7 (100\%) & \colorbox{rateHighBlue}{\textbf{\textcolor{white}{6 (86\%)}}} & 5 (71\%) & 2 (40\%) & 5 (71\%) & 0 (0\%) & 5 (71\%) & \colorbox{rateHighBlue}{\textbf{\textcolor{white}{5 (100\%)}}} \\
 & number & 2 (29\%) & \colorbox{rateHighBlue}{\textbf{\textcolor{white}{2 (100\%)}}} & 6 (86\%) & \colorbox{rateHighBlue}{\textbf{\textcolor{white}{6 (100\%)}}} & 6 (86\%) & \colorbox{rateHighBlue}{\textbf{\textcolor{white}{6 (100\%)}}} & 7 (100\%) & \colorbox{rateHighBlue}{\textbf{\textcolor{white}{7 (100\%)}}} & 7 (100\%) & \colorbox{rateHighBlue}{\textbf{\textcolor{white}{7 (100\%)}}} \\ \hline
 
\multirow{2}{*}{Total} & string & 44 (48\%) & 30 (68\%) & 89 (98\%) & \colorbox{rateHighBlue}{\textbf{\textcolor{white}{74 (83\%)}}} & 84 (92\%) & 7 (8\%) & 78 (86\%) & 6 (8\%) & 87 (96\%) & \colorbox{rateHighBlue}{\textbf{\textcolor{white}{78 (90\%)}}} \\
 & number & 12 (50\%) & \colorbox{rateHighBlue}{\textbf{\textcolor{white}{11 (92\%)}}} & 21 (88\%) & \colorbox{rateHighBlue}{\textbf{\textcolor{white}{21 (100\%)}}} & 17 (71\%) & \colorbox{rateHighBlue}{\textbf{\textcolor{white}{17 (100\%)}}} & 15 (62\%) & \colorbox{rateHighBlue}{\textbf{\textcolor{white}{15 (100\%)}}} & 17 (71\%) & \colorbox{rateHighBlue}{\textbf{\textcolor{white}{17 (100\%)}}} \\ \hline
 
\end{tabularx}
\end{adjustbox}
\end{table*}
\begin{table*}
\centering
\scriptsize
\caption{Judge summary group by project with all string contexts}
\label{tab:judge_all}
\begin{tabular}{c|c|cc|cc|cc|cc|cc}
 \hline
\multicolumn{2}{c|}{\multirow{2}{*}{\textbf{Project}}} & \multicolumn{2}{c|}{\textbf{Manual}} & \multicolumn{2}{c|}{\textbf{ChatUniTest}} & \multicolumn{2}{c|}{\textbf{EvoSuite}} & \multicolumn{2}{c|}{\textbf{Randoop}} & \multicolumn{2}{c}{\textbf{EvoSuiteC3}}  \\ \cline{3-12} 
\multicolumn{2}{c|}{} & \textbf{Cover. (\%)} & \textbf{Read. (\%)} & \textbf{Cover. (\%)} & \textbf{Read. (\%)} & \textbf{Cover. (\%)} & \textbf{Read. (\%)} & \textbf{Cover. (\%)} & \textbf{Read. (\%)}   & \textbf{Cover. (\%)} & \textbf{Read. (\%)}  \\ \hline

\multirow{1}{*}{Total} & string & 50 (48\%) & 31 (62\%) & 102 (98\%) & \colorbox{rateHighBlue}{\textbf{\textcolor{white}{76 (75\%)}}} & 96 (92\%) & 7 (7\%) & 90 (87\%) & 6 (7\%) & 100 (96\%) & \colorbox{rateHighBlue}{\textbf{\textcolor{white}{91 (91\%)}}}\\ \hline
 
\end{tabular}
\end{table*}
\begin{table*}
\centering
\caption{Average judge summary of all tests from EvoSuite, Randoop, and EvoSuiteC3 with true positive contexts}
\label{tab:judge30}
\begin{tabular}{c|c|cc|cc|cc}
 \hline
\multicolumn{2}{c|}{\multirow{2}{*}{\textbf{Project}}} & \multicolumn{2}{c|}{\textbf{EvoSuite}} & \multicolumn{2}{c|}{\textbf{Randoop}} & \multicolumn{2}{c}{\textbf{EvoSuiteC3}} \\ \cline{3-8} 
\multicolumn{2}{c|}{} & \textbf{Cover. (\%)} & \textbf{Read. (\%)} & \textbf{Cover. (\%)} & \textbf{Read. (\%)} & \textbf{Cover. (\%)} & \textbf{Read. (\%)}  \\ \hline

\multirow{2}{*}{Airsonic} & string  & 20.25 (96\%) & 3.54 (17\%) & 20.9 (100\%) & 5.67 (27\%)  & 21.0 (100\%) & \colorbox{rateHighBlue}{\textbf{\textcolor{white}{20.15 (96\%)}}} \\
 & number  & 9.7 (61\%) & \colorbox{rateHighBlue}{\textbf{\textcolor{white}{9.65 (99\%)}}} & 8.43 (53\%) & \colorbox{rateHighBlue}{\textbf{\textcolor{white}{8.4 (100\%)}}}  & 9.92 (62\%) & \colorbox{rateHighBlue}{\textbf{\textcolor{white}{9.92 (100\%)}}} \\ \hline
 
\multirow{2}{*}{GoogleMapService} & string  & 15.36 (81\%) & 0.69 (4\%) & 13.47 (71\%) & 0.23 (2\%)  & 16.28 (86\%) & 10.77 (66\%) \\
 & number  & - & - & - & -  & - & - \\ \hline
 
\multirow{2}{*}{OpenMRS} & string  & 25.5 (98\%) & 1.83 (7\%) & 22.37 (86\%) & 1.4 (6\%)  & 26.0 (100\%) & \colorbox{rateHighBlue}{\textbf{\textcolor{white}{23.33 (90\%)}}} \\
 & number  & - & - & - & -  & - & - \\ \hline
 
\multirow{2}{*}{Shopizer} & string  & 17.94 (100\%) & 1.84 (10\%) & 18.0 (100\%) & 1.5 (8\%)  & 18.0 (100\%) & \colorbox{rateHighBlue}{\textbf{\textcolor{white}{17.8 (99\%)}}} \\
 & number  & 0.0 (0\%) & - & 0.0 (0\%) & -  & 0.0 (0\%) & - \\ \hline
 
\multirow{2}{*}{Twitter4J} & string  & 5.0 (71\%) & 3.13 (63\%) & 5.97 (85\%) & 3.53 (59\%)  & 5.0 (71\%) & \colorbox{rateHighBlue}{\textbf{\textcolor{white}{4.97 (99\%)}}} \\
 & number  & 6.43 (92\%) & \colorbox{rateHighBlue}{\textbf{\textcolor{white}{6.43 (100\%)}}} & 6.53 (93\%) & \colorbox{rateHighBlue}{\textbf{\textcolor{white}{6.53 (100\%)}}}  & 6.43 (92\%) & \colorbox{rateHighBlue}{\textbf{\textcolor{white}{6.43 (100\%)}}} \\ \hline
 
\multirow{2}{*}{Total} & string  & 84.05 (92\%) & 11.03 (13\%) & 80.71 (88\%) & 12.33 (15\%)  & 86.28 (94\%) & \colorbox{rateHighBlue}{\textbf{\textcolor{white}{77.02 (89\%)}}} \\
 & number  & 16.13 (67\%) & \colorbox{rateHighBlue}{\textbf{\textcolor{white}{16.08 (99\%)}}} & 14.96 (62\%) & \colorbox{rateHighBlue}{\textbf{\textcolor{white}{14.93 (99\%)}}}  & 16.35 (68\%) & \colorbox{rateHighBlue}{\textbf{\textcolor{white}{16.35 (100\%)}}} \\ \hline
 
\end{tabular}
\end{table*}
\subsection{RQ2: Readability of Manual and Automated Tests}\label{sec:eval:compare}

This RQ aims to study the differences between manual and automated tests under C3 measurement.

\subsubsection{Methodology}

\noindent\textit{(1) Subjects.}
Table~\ref{tab:projects_rq2} shows this RQ's subjects, i.e., $30$ classes with manual tests whose false positive cases have been identified in RQ1 (see Sec.~\ref{sec:eval:accurate}). These $30$ classes contain $128$ parameters with context detected by C3, of which $115$ were manually confirmed as true positive cases. We chose these classes because they have manual tests, and we aim to compare manual and automated tests in this RQ.

\noindent\textit{(2) Baselines.}
We have four baselines: manual tests, EvoSuite~\cite{FraserEvoSuite}, Randoop~\cite{pacheco2007randoop}, ChatUniTest~\cite{xie2023chatunitest}. We use EvoSuite and Randoop to represent two traditional test generation techniques, i.e., search-based software testing~\cite{FraserWhole, FraserEvoSuite, PanichellaMOSA, PanichellaDynaMOSA, LinGraph} and random software testing~\cite{pacheco2007randoop, PachecoFeedback, pacheco2005eclat, andrews2011genetic}. We use ChatUniTest to represent LLM-based test generation~\cite{tufano2020unit, xie2023chatunitest, wang2023software} since it is ready-to-use and designed to continuously guide LLM in generating syntactically correct tests without manual intervention.

\noindent\textit{(3) Test generation procedure.}
For EvoSuite and Randoop, we use their default settings to run on each class for producing tests. Due to their randomness, we repeat the procedure for $30$ rounds, and each round's time budget is two minutes. For ChatUniTest, we adopt the authors' settings~\cite{xie2023chatunitest}. For example, the test number for each method under test (MUT) is six, and the LLM is GPT-3.5-Turbo~\cite{GPTModels}. The only difference is that we use at most ten rounds of calling GPT-3.5-Turbo to generate and repair a test instead of six rounds to obtain more syntactically correct tests. Although GPT-3.5-Turbo also has randomness, we only execute the routine once due to financial costs.

\noindent\textit{(4) C3 measurement procedure.}
After the test generation procedure, we employ C3's judging component (see Sec.~\ref{subsec:ccc_judge}) to evaluate all tests. We execute C3 with GPT-4-Turbo on both manual and automated tests produced by ChatUniTest. Likewise, we utilize GPT-4-Turbo for the first round of tests generated by EvoSuite and Randoop. However, considering the substantial model financial costs brought by running GPT-4-Turbo for the tests of $30$ rounds, we use GPT-3.5-Turbo to run with C3 for them since GPT-3.5-Turbo is much cheaper than GPT-4-Turbo.

\noindent\textit{(5) Data analysis.}
We calculate two metrics for both manual and automated tests generated by ChatUniTest, as well as the first round of EvoSuite and Randoop tests: the number of covered parameters (i.e., the existence of a test that calls the MUT with an input) and parameters with readable inputs (i.e., the existence of a test that calls the MUT with a \textbf{readable} input). Additionally, we show the metric data by true positive contexts and all contexts (see Table~\ref{tab:projects_rq2}), respectively, to show the effects of false positive cases from C3 on measuring test generation tools. Note that we only show string-type data for all contexts since there are no false positive cases in number-type contexts~(see Table~\ref{tab:projects_rq2}). Furthermore, we compute the average values of these two metrics across all rounds of tests generated by EvoSuite and Randoop to check whether the randomness affects their performance in generating readable inputs.

\subsubsection{Results}

\noindent\textit{(1) String coverage and readability.}
Table~\ref{tab:judge_tp} presents the performance of four baselines (and EvoSuiteC3, see Sec.~\ref{sec:eval:evoc3}) in covered parameters and parameters with readable inputs on $91$ true positive string-type contexts. Overall, three automated tools surpass manual tests in parameter coverage ($86\%$ to $98\%$ versus $48\%$). The readability rates (i.e., the ratio of parameters with readable input to covered parameters) indicate that ChatUniTest and manual tests significantly outperform EvoSuite and Randoop in generating readable inputs ($83\%$ and $68\%$ versus $8\%$ and $8\%$). Fig.~\ref{fig:wordcloud} summarizes test inputs of four baselines as well as EvoSuiteC3.

\noindent\textit{(2) Number coverage and readability.}
Table~\ref{tab:judge_tp} also presents baselines' performance on $24$ number-type contexts. Similarly, three test generation tools (ChatUniTest, EvoSuite, and Randoop) surpass manual tests in parameter coverage ($88\%$, $71\%$, and $62\%$ versus $50\%$). Their readability rates are close: $92\%$ (manual tests) and $100\%$ (three tools).

\noindent\textit{(3) Effects of false positive contexts.}
Table~\ref{tab:judge_all} shows the results of all string-type contexts (including false positive cases). The parameter coverage delta between all and true positive contexts only exists in Randoop ($86\%$ and $87\%$). The readability deltas range from $1\%$ (EvoSuite) to $8\%$ (ChatUniTest).

\noindent\textit{(4) Thirty rounds.}
We run EvoSuite and Randoop $30$ times to eliminate the randomness. Table~\ref{tab:judge30} shows their average performance. In total, the parameter coverage deltas between one and $30$ rounds range from $0\%$ (EvoSuite's string) to $4\%$ (EvoSuite's number). The readability deltas range from $1\%$ (EvoSuite's number) to $7\%$ (Randoop's string).

\subsubsection{Discussion}

\noindent\textit{(1) Parameter coverage.}
The parameter coverage figures show that all test generation tools generate higher code coverage than manual tests, confirming many studies' claims and findings~\cite{Rojas2015CombiningMC, wang2021ML}.

\noindent\textit{(2) String readability.}
ChatUniTest's tests surpass manual ones regarding string-input readability, while EvoSuite and Randoop almost fail to generate readable string-type inputs.

\noindent\textit{(3) Number readability.}
All baselines achieve scores near $100\%$ in number readability. This is primarily because all these contexts are \enquote{LONGNUMBER}, and C3 regards that a number shorter than $9$ meets this context's readability (see Sec.~\ref{subsec:ccc_judge}). Meanwhile, such numbers are prevalent in all baseline tests.

\noindent\textit{(4) Effects of false positive contexts.}
False positive contexts lead to underestimating the string-type readability performance of ChatUniTest and manual tests. The primary reason is that these two baselines perform well in string-type readability, thus making them sensitive to incorrect contexts. However, such contexts have a tiny impact on EvoSuite and Randoop, as these tools fail to generate readable string inputs. Therefore, C3, even with false positive contexts, is sufficient to evaluate traditional tools.

\noindent\textit{(5) Thirty rounds.}
The thirty rounds of data confirm that EvoSuite and Randoop lag far behind manual tests and ChatUniTest in generating readable string inputs.

\begin{tcolorbox}[title=Answer to RQ2,boxrule=1pt,boxsep=1pt,left=2pt,right=2pt,top=2pt,bottom=2pt]
ChatUniTest's tests surpass manual ones regarding parameter coverage and C3 measurement, while EvoSuite and Randoop almost fail to generate readable string-type inputs.
\end{tcolorbox}
\subsection{RQ3: EvoSuiteC3's Improvement}\label{sec:eval:evoc3}
Sec.~\ref{sec:eec} introduces EvoSuiteC3, i.e., EvoSuite enhanced by C3. In this RQ, we aim to determine how much improvement EvoSuiteC3 brings to EvoSuite.
\subsubsection{Methodology}
\noindent\textit{(1) Subjects.}
We use the same subjects as RQ2 (shown in Table~\ref{tab:projects_rq2}).

\noindent\textit{(2) Baselines.}
We mainly use EvoSuite to compare with EvoSuiteC3 to evaluate the improvements brought by EvoSuiteC3. We also use ChatuniTest and manual tests as baselines to compare EvoSuiteC3 since C3 confirmed their excellent input readability in RQ2 (see Sec.~\ref{sec:eval:compare}).

\noindent\textit{(3) Test generation procedure.}
EvoSuiteC3 is based on EvoSuite and does not introduce any extra running parameters. Hence, we use EvoSuite's default settings to run EvoSuiteC3 on each class for producing tests. Due to its randomness, we repeat the procedure for $30$ rounds to generate $30$ test suites, and each round's time budget is two minutes. Besides, we pass all contexts, including false positive ones mined by C3 (Table~\ref{tab:projects_rq2}), to EvoSuiteC3.

\noindent\textit{(4) C3 measurement procedure.}
After the test generation, we use C3's judging component (see Sec.~\ref{subsec:ccc_judge}) to evaluate EvoSuiteC3's tests. Like in RQ2, we run C3 with GPT-4-Turbo for the first round of tests generated by EvoSuiteC3. Meanwhile, we use GPT-3.5-Turbo for all rounds of tests since using GPT-4-Turbo for all rounds of tests would cost significant financial costs, and GPT-3.5-Turbo is much cheaper than GPT-4-Turbo.

\noindent\textit{(5) Data analysis.}
We analyze EvoSuiteC3 tests in the same way as we analyzed EvoSuite tests in RQ2 (see Sec.~\ref{subsec:ccc_judge}), i.e., we focus on the number of covered parameters and parameters with readable inputs. Furthermore, EvoSuiteC3 defines two kinds of fitness functions for EvoSuite (see Sec.~\ref{sec:eec}). First, $f_{c3}$ measures whether an input satisfies a parameter's context. Second, $f_{c3invo}$ measures whether a method invocation's inputs satisfy all contexts of this method's parameters. After the test generation, we recorded the ratios of the covered fitness functions (i.e., the fitness functions reaching the optimum) to all functions for each generated test suite as two metrics, C3 Coverage and \ctinvo{} Coverage.
\subsubsection{Results}

\noindent\textit{(1) Comparison with \evosuite{}.}
The last column of Tables~\ref{tab:judge_tp},~\ref{tab:judge_all}, and~\ref{tab:judge30} shows the C3 measurement data of EvoSuiteC3. They uniformly show that with other metrics almost the same, the string readability increases from about $10\%$ to around $90\%$. Code~\ref{code:airsonic_user} presents an example. EvoSuite generates two random inputs for two parameters (\inlinecode{username} and \inlinecode{email}), while EvoSuiteC3 generates a person's name and an email address for them. Fig.~\ref{fig:wordcloud} confirms that EvoSuiteC3 generates readable inputs while EvoSuite's inputs are commonly random.

\noindent\textit{(2) Comparison with Manual Tests and ChatUniTest.}
Table~\ref{tab:judge_tp} presents the C3 measurement data of manual tests, ChatUniTest, and EvoSuiteC3. First, ChatUniTest and EvoSuiteC3 surpass manual tests in both parameter coverage and string readability. Second, EvoSuiteC3 reaches the highest string readability for all classes ($90\%$). EvoSuiteC3 only performs worse than manual tests in the GoogleMapService project's string readability score ($76\%$ versus $83\%$).

\noindent\textit{(3) C3 Coverage and \ctinvo{} Coverage.}
Table~\ref{table:c3coverage} shows the mean and median coverage data of EvoSuiteC3. Since $f_{c3invo}$ measures whether a method invocation's inputs satisfy all contexts, we regard that the number of parameters with context that a method contains significantly impacts its performance of guiding genetic algorithms. Hence, we separately list classes containing methods with at least one, two, and three parameters with contexts to show the coverage data. The mean and median of C3 Coverage remain stable at above $90\%$ and $100\%$ under different class groups, respectively. Conversely, \ctinvo{} Coverage decreases as the number of parameters with contexts of a method increases.

\begin{table}[htbp]
\centering
\begin{tabular}{c|c|c|c|c|c}
\hline
\multirow{2}{*}{Class Selector} & \multirow{2}{*}{Classes} & \multicolumn{2}{c|}{C3 Coverage} & \multicolumn{2}{c}{\ctinvo{} Coverage} \\ \cline{3-6}
                   &                         & Mean      & Median    & Mean      & Median    \\ \hline
1                       & 30                      & 93.5\%    & 100\%     & 89.1\%    & 100\%     \\ \cline{3-6} \hline
2                       & 7                       & 94\%      & 100\%     & 78\%      & 75\%      \\ \cline{3-6} \hline
3                       & 3                       & 99.8\%    & 100\%     & 73.7\%    & 75\%      \\ \cline{3-6} \hline
\end{tabular}
\caption{Overview of C3 Coverage and \ctinvo{} Coverage: Class Selector selects classes that have methods containing $\geq$ N parameters with contexts}
\label{table:c3coverage}
\end{table}
\subsubsection{Discussion}
First, C3 measurement data show that EvoSuiteC3 significantly improves EvoSuite's string readability, confirming that our fitness functions successfully help EvoSuite generate readable inputs. Second, Table~\ref{tab:judge_tp} shows that EvoSuiteC3 only performs worse than manual tests in the GoogleMapService project's string readability score. The main reason is the manual tests' lower parameter coverage. They only cover $12$ parameters, and $10$ ($83\%$) have readable inputs, but the EvoSuiteC3 tests cover $17$ parameters, and $13$ ($76\%$) have readable inputs. Third, the readability scores under C3 measurement and the ratios of covered $f_{c3}$ fitness functions to all $f_{c3}$ fitness functions (C3 Coverage) are close (around $90\%$ to $100\%$). These data confirm that our fitness functions successfully reflect the readability requirements of C3. Fourth, since we pass all contexts to EvoSuiteC3, its performance is nearly identical with or without false positive contexts. Besides, \ctinvo{} Coverage decreases as parameters with contexts of a method increase. We leverage the \inlinecode{User} class of the Airsonic project shown in Code~\ref{code:airsonic_user} as an example. Its constructor method contains one parameter with the \enquote{PERSON} context and another with the \enquote{EMAIL} context. Among the $30$ test suites generated by EvoSuiteC3, all suites contain a test satisfying the \enquote{PERSON} context and a test satisfying the \enquote{EMAIL} context. Conversely, only $13$ suites contain a test satisfying both the \enquote{PERSON} and \enquote{EMAIL} context. The results show that our $f_{c3invo}$ fitness functions are not effective enough. The main reason is that $f_{c3invo}$ combines multiple $f_{c3}$ functions and returns the max of their return values (see Equation~\ref{fitness:c3invo}). $f_{c3invo}$ is satisfied if only all its $f_{c3}$ are satisfied. Hence, it is more difficult for genetic algorithms to generate tests satisfying the $f_{c3invo}$ functions. We plan to investigate this issue in the future.
\begin{tcolorbox}[title=Answer to RQ3,boxrule=1pt,boxsep=1pt,left=2pt,right=2pt,top=2pt,bottom=2pt]
Under C3 measurement, EvoSuiteC3 significantly improves EvoSuite's string input readability by satisfying from about $10\%$ of contexts to around $90\%$.
\end{tcolorbox}

\subsection{RQ4: Programmers' Readability Measurements}\label{sec:eval:survey}
In this RQ, we aim to determine whether programmers agree with C3's measurement of tests' readability.

\subsubsection{Methodology}
\begin{figure}[t!]
    \centering
    \includegraphics[width=0.46\textwidth]{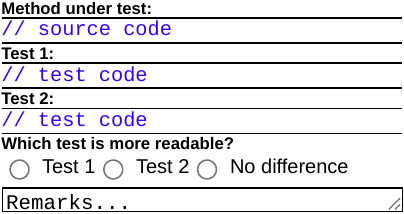}
    \caption{The question framework of the questionnaire}
    \label{fig:form}
\end{figure}
\noindent\textit{(1) Subjects.}
From manual and automated tests used in Sec. \ref{sec:eval:compare}, we randomly select ten pairs, each testing the same method and covering identical readability contexts. Each pair consists of a manual and an automated test. Of these ten pairs, C3 identifies manual tests of five pairs as more readable than automated ones (referred to as C3-Diff pairs). In comparison, the other five pairs are determined to have equivalent readability (C3-Equal pairs).

\noindent\textit{(2) Questionnaire.} 
We arrange these ten pairs into a questionnaire, forming ten questions by applying the question framework depicted in Fig.~\ref{fig:form} to each pair of tests. We shuffle a pair's tests to ensure that a manual or an automated test is not always shown in the \textbf{Test 1} or \textbf{Test 2} of the framework. Note that we aim to determine whether C3 reflects programmers' requirements for test readability. Hence, the questionnaire only asked interviewees which test was more readable because we did not want to inform respondents explicitly about differences in test inputs. Otherwise, it could cause respondents to focus too much on the inputs, influencing their judgments about the tests' readability. After completing the questionnaire, we actively communicated with the respondents about the impact of the inputs on readability.

\noindent\textit{(3) Participants.}
Following the previous study~\cite{panichella2016impact}, we invite $30$ participants to answer the questionnaire. Table~\ref{tab:participant_information} shows their jobs and programming experiences.
\begin{table}[htbp]
  \centering
  \caption{Participants' jobs and programming experiences}
  \begin{minipage}[t]{.5\linewidth}
  \vspace{0pt}
  \centering
    \begin{tabular}{c|c}
    \hline
    \textbf{Job} & \textbf{Number} \\
    \hline
    Developer & 17 \\
    Graduate Student & 11 \\
    Professor & 1 \\
    Test Engineer & 1 \\
    \hline
    \end{tabular}%
  \label{tab:addlabel}%
  \end{minipage}%
  \begin{minipage}[t]{.5\linewidth}
  \vspace{0pt}
  \centering
    \begin{tabular}{c|c}
    \hline
    \textbf{Experience} & \textbf{Number} \\
    \hline
    Over 3 years & 27 \\
    1 to 3 years & 2 \\
    Within 1 year & 1 \\
    \hline
    \end{tabular}%
  \label{tab:participant_information}%
  \end{minipage}%
\end{table}%

\noindent\textit{(4) Data analysis.}
After participants finish the questionnaire, firstly, we group the data by C3-Diff and C3-Equal, then decode and count their choices, which include \enquote{manual is better}, \enquote{automated is better}, and \enquote{no difference}. Secondly, for each group, we calculate the Cohen's Kappa coefficient~\cite{cohen1968weighted, landis1977measurement, mchugh2012interrater} for every two participants to determine their agreement level. The Cohen's Kappa coefficient, ranging from $-1$ to $1$, measures the agreement between two raters, where $1$ means perfect agreement, $0.61$ means substantial agreement, and $0$ means random agreement~\cite{landis1977measurement, mchugh2012interrater}.

\subsubsection{Results}
\begin{table}[h]
\centering
\caption{Participants' choice summary}
\begin{tabular}{c|c|c|c|c}
\hline
\textbf{Group} & \textbf{Manual} & \textbf{Automated} & \textbf{No Diff.} & \textbf{Total} \\
\hline
C3-Equal & 52 & 88 & 10 & 150 \\
\hline
C3-Diff & 88 & 47 & 15 & 150 \\
\hline
\end{tabular}
\label{tab:participant_choice}%
\end{table}
\begin{figure}[t!]
    \centering
    \includegraphics[width=0.47\textwidth]{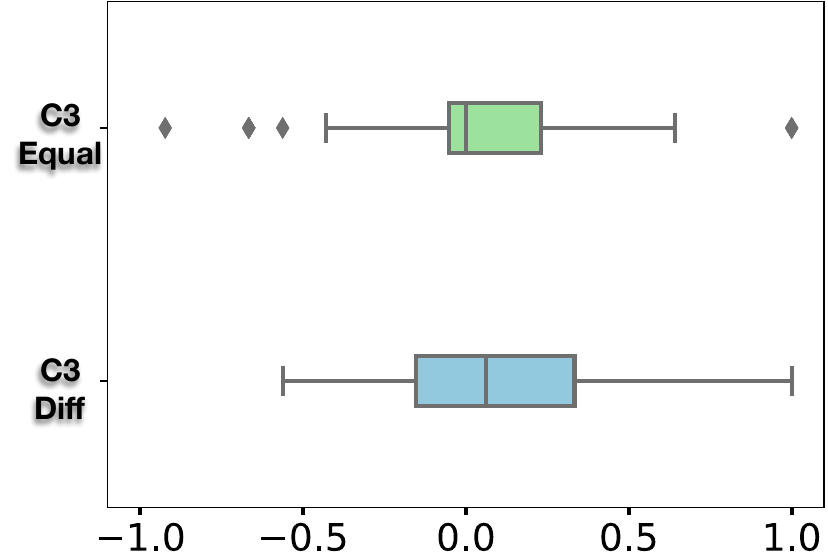}
    \caption{Participants' Cohen's Kappa coefficients}
    \label{fig:ck}
\end{figure}

\noindent\textit{(1) Choice summary.}
Table~\ref{tab:participant_choice} summarizes participants' choices. When C3 asserts the manual test is more readable, 88 out of 150 instances ($58.7\%$) show agreement, while 47 instances ($31.3\%$) display explicit disagreement. For the C3-Equal cases, explicit preferences are expressed in $52+88=140$ instances ($93.3\%$).

\noindent\textit{(2) Cohen's Kappa coefficients.}
Fig.~\ref{fig:ck} displays the levels of agreement among participants, grouped by C3-Diff and C3-Equal. The agreement levels for C3-Diff range from $-0.563$ to $1$, whereas for C3-Equal, they range from $-0.923$ to $1$. The median values for C3-Diff and C3-Equal are $0.063$ and $0$, respectively, while the mean values are $0.118$ and $0.067$. For C3-Diff, there are $37$ instances reaching substantial agreement (values $\geq 0.61$), compared to $23$ such instances for C3-Equal.

\subsubsection{Discussion}
We observe that for C3-Diff pairs, where C3 deems manual tests more readable than automated ones, $103$ instances ($68.7\%$) of participants' choices express at least no objection. Moreover, several metrics of Cohen's Kappa coefficients from C3-Diff pairs surpass those from C3-Equal pairs, including the minimum, median, and mean values and substantial agreement count. These findings suggest that when C3 distinguishes readable differences between two tests, programmers are more inclined to reach a consensus and concur with C3. However, $31.3\%$ of choices still disagree with C3 in C3-Diff pairs. Concurrently, the choice summary and Cohen's Kappa coefficients of C3-Equal pairs indicate that programmers' preferences are not uniform. This is because apart from test inputs matching contexts, different programmers emphasize various other factors in readability, which are shown in the programmers' feedback.

\noindent\textbf{Feedback from programmers.}
After completing the questionnaire, we specifically asked participants about the importance of matching test inputs to contexts for readability and which other factors they emphasize. We categorize their feedback as follows:

\noindent\textit{(1) Programmers approve readability context matching.}
Most programmers concur that matching test inputs to contexts enhances readability. Twelve participants expressed their dissatisfaction with random and meaningless inputs (generated by EvoSuite and Randoop). One participant stated that the situation deteriorates when a test calls a method with multiple random inputs, as the mapping from values to parameters becomes unidentifiable. The EvoSuite test of Code~\ref{code:airsonic_user} provides an example.
\begin{javacode}{Two object creation statements}{code:s_e}
// object creation of manual test
OEmbed oembed = twitter1.getOEmbed(new OEmbedRequest(240192632003911681L, "http://somesite.com/"));
// object creation of automated test
OEmbedRequest oEmbedRequest0 = new OEmbedRequest(1784L, "kP");
\end{javacode}

\noindent\textit{(2) Short and explicit tests are preferred.}
Beyond input contexts, two hot keywords from programmers discussing test readability are shortness and explicitness. Shortness is a requirement for all test elements, such as variable names, statements, and literals. Explicitness implies that (1) a test should be self-contained, i.e., the overuse of methods other than the MUT is discouraged. This prevents readers from referring to different methods for understanding the test; (2) statements should be straightforward. For instance, pass-by-value is preferable to pass-by-variable when calling the MUT. Code~\ref{code:s_e} exemplifies this: the manual test violates both principles (too long literals and calling two methods), whereas the automated one adheres to them. Hence, $21$ out of $30$ programmers preferred the latter, although the manual one satisfies the URL context. However, for the MUT in Code~\ref{code:airsonic_user} with two parameters requiring readability contexts, $22$ out of $30$ preferred the manual test to the EvoSuite test, indicating dominating factors vary per scenario.

\noindent\textit{(3) Unreadable inputs are necessary.}
Two participants mentioned that unreadable inputs are necessary when testing boundary conditions since many defects are triggered by unexpected inputs. This posed a challenge for C3 to measure tests since C3 could not identify unreadable inputs on purpose. Still, our experimental setup avoided this issue: For a context, as long as one input meets the requirement, we assert that context is to be satisfied (see Sec.~\ref{sec:eval:compare}).

\begin{tcolorbox}[title=Answer to RQ4,boxrule=1pt,boxsep=1pt,left=2pt,right=2pt,top=2pt,bottom=2pt]
When C3 identifies readable differences between tests, programmers tend to reach a consensus and agree with it, although C3 alone does not include all readability factors.
\end{tcolorbox} 

\section{Case Study: Adding a new Context}\label{sec:addcontext}
In C3's Mining Component (Sec.~\ref{subsec:ccc_mine}), we prepare a group of contexts to restrict LLMs to choosing one of them or answering no context for parameters. Such design choice prevents LLMs from answering open-ended questions, thus mitigating LLMs' hallucinations~\cite{fan2023large, huang2024visual}. However, our prepared contexts might not cover all the contexts of real-world projects. To cope with this issue, we allow users to add new contexts to C3. We leverage an example for our evaluation (Sec.~\ref{sec:eval:accurate}) to show how to add new contexts.

This example comes from OpenMRS~\cite{Openmrs} (an application recording medical events). It contains a string parameter, \textbf{\inlinecode{role}}. This parameter requires readable inputs, and some instances include ``Medical Student'', ``Doctor'', ``Data Manager'', etc. However, it was not recognized with a context in our evaluation since we did not define MRS's Role context in Sec.~\ref{subsec:ccc_mine}. After adding the context name (\enquote{MRS\_ROLE}) and some input examples like the above-mentioned instances, C3's prompt for LLMs to mine contexts (see Fig.~\ref{fig:mining_example}) added one more option for LLMs' response. We used a concrete LLM (GPT-4-Turbo) to test C3, and C3 successfully recognized this parameter with the ''MRS\_ROLE`` context.

Meanwhile, we prepared two tests shown in Code~\ref{code:role}. Using GPT-4-Turbo, C3's Judging Component (Sec.~\ref{subsec:ccc_judge}) also gave the correct result that ``Nurse'' is readable while ``xxx'' is not. ``Nurse'' is not one of the predefined input examples, confirming LLMs' flexibility in the judging task. Note that in Sec.~\ref{subsec:ccc_judge}, we also use traditional NLP tools (e.g., CoreNLP~\cite{manning2014stanford}) to judge input readability via the named entity recognition (NER) function. However, such tools could not be applied to customized contexts like \enquote{MRS\_ROLE} without manual extension. Conversely, LLMs' support for arbitrary inputs allows us to support custom contexts easily.
\begin{javacode}{Two tests for the role parameter}{code:role}
void test1()  {
    Role role = new Role("xxx"); 
}
void test2()  {
    Role role = new Role("Nurse"); 
}
\end{javacode}
\section{Threats to Validity}\label{sec:threats}

\noindent\textbf{Construct Validity.}
In Sec.~\ref{subsec:ccc_mine}, we use $23$ named entities~\cite{sang2003introduction, ritter2011named, manning2014stanford, wang2023gpt} to construct string type' contexts, which brings a threat that such named entities are not complete to represent all contexts of real-world projects. We provide an example in Sec.~\ref{sec:eval:accurate}. Firstly, We plan to extend C3 to mine contexts beyond prepared contexts in the source code automatically. Secondly, in practice, we allow C3's users to integrate new contexts into C3 (see Sec.~\ref{sec:addcontext}).

\noindent\textbf{Internal Validity.}
The first threat to internal validity comes from two parameters (\textit{MaxToken} and \textit{Remain}) of C3's Mining Component (Sec.~\ref{subsec:ccc_mine}). We use \textit{MaxToken} to represent the token size limit and \textit{Remain} to keep the token space for LLM's response. With these two parameters, our prompt token size will not exceed the LLM's token limit. In our experiment, we used GPT-3.5/4-turbo, and their token limit was $4096$ when we ran our experiment. C3' users need to reset \textit{MaxToken} when their adopted LLM has another token limit. \textit{Remain} is used to remain the token space for LLM's response. C3's expected response should be a small string, i.e., a context name (e.g., ``EMAIL'' and ``PERSON''). Hence, we set \textit{Remain} to a small number ($20$ in our experiment). However, C3's users also need to adjust this parameter's concrete value when using an LLM with a different string-tokenization algorithm from GPT-3.5/4 since a different tokenization algorithm may lead to a significantly different token size, even for a small string.  The second threat to internal validity comes from a parameter (\textit{BearableLength}) of the C3's Judging Component (Sec.~\ref{subsec:ccc_judge}). This parameter is used to control the readable threshold for Number's three contexts (FIXEDLENGTH, LONGNUMBER, and SCIENTIFIC). For these three contexts, when the inputs are not too long (e.g., $12$ and $12345$), we regard them as readable. Hence, we add the bearable length to filter out such numbers when we apply our judger on the tests. Since many meaningful numbers (e.g., American phone numbers and credit card numbers) usually have at least ten digits, we set the bearable length at $9$. The third threat is the randomness of the test generation tools (ChatUnitTest, EvoSuiteC3, EvoSuite, and Randoop) when comparing their test input readability in Sec.~\ref{sec:eval:compare}. To reduce this risk, We repeated running EvoSuiteC3, EvoSuite, and Randoop $30$ times on the same \java{} class. However, we only ran ChatUnitTest once because it leveraged GPT-3.5-turbo to generate tests and could bring many financial costs if we repeated running it $30$ times. The last threat is due to potential bugs introduced in implementing C3 and EvoSuiteC3. For C3, we manually checked its extracted contexts and confirmed that its Precision, Recall, and F1-Score are \precision{}, \recall{}, and \fone{}, respectively (see Sec.~\ref{sec:eval:accurate}). For EvoSuiteC3, we extracted the inputs from its generated tests and manually checked whether these inputs match the contexts. The result confirmed EvoSuite's readability increase brought by EvoSuiteC3. Fig.~\ref{fig:wordcloud} demonstrates the comparison between EvoSuiteC3 and EvoSuite.

\noindent\textbf{External Validity.}
Experimental subjects form a threat. Many classes from the DynaMOSA benchmark~\cite{PanichellaDynaMOSA} are obsolete~\cite{LinGraph}. Hence, we complement it by adding $155$ classes from seven open-source projects (see Sec.~\ref{sec:eval:accurate}). The choice of LLMs in our evaluation forms another threat since C3's performance depends on LLMs. Moreover, we implemented C3 in \java{} and evaluated it using \java{} subjects. Hence, our experimental findings are not verified in other programming languages. Nonetheless, the fundamental concepts of C3 are not confined to \java{}.
\section{Conclusion}\label{sec:conclude}
This paper proposes the Context Consistency Criterion (C3), an automatic readability measurement tool for primitive-type test input. C3 mines the readability contexts for primitive-type parameters in the code under test and uses these contexts to judge tests' consistency as a readability metric. Besides, this paper proposes EvoSuiteC3 (i.e., EvoSuite enhanced by C3). Our experimental results show the distribution of readability contexts and confirm C3's accuracy. Furthermore, while LLM-based test generation and EvoSuiteC3 surpass manual tests under C3 measurement, two traditional tools (Randoop and the original EvoSuite) almost fail to generate readable string-type inputs.

\section{Data Availability}\label{sec:data}
C3's implementation and the experimental data are at \url{https://doi.org/10.5281/zenodo.13254261}.
\section*{Acknowledgements}
This paper is partially sponsored by the Shanghai Sailing Program (No. 22YF1428600) and the National Natural Science Foundation of China (No. 62202306).
\ifCLASSOPTIONcaptionsoff
\newpage
\fi
\bibliographystyle{IEEEtran}
\bibliography{ref}
\balance

\end{document}